%% file: acm.tex
\documentclass[sigconf]{acmart}

\usepackage{algorithmic}
\usepackage{graphicx}
\usepackage{textcomp}
\usepackage{xcolor}
\usepackage[utf8]{inputenc} 
\usepackage[T1]{fontenc}    

\usepackage{url}            
\usepackage{booktabs}       
\usepackage{amsfonts}       
\usepackage{nicefrac}       
\usepackage{microtype}      
\usepackage{xcolor}         
\usepackage{color}
\usepackage{paralist}
\usepackage{amsmath}
\usepackage{graphicx}
\usepackage{subfigure}
\usepackage{array}
\usepackage{epstopdf}
\usepackage[ruled,linesnumbered]{algorithm2e}
\usepackage{multirow} 
\usepackage{tabularx}

\usepackage{changes}
\usepackage{threeparttable}
\usepackage{tikz}
\newcommand*\emptycirc[1][1ex]{\tikz\draw (0,0) circle (#1);} 
\newcommand*\halfcirc[1][1ex]{%
	\begin{tikzpicture}
	\draw[fill] (0,0)-- (90:#1) arc (90:270:#1) -- cycle ;
	\draw (0,0) circle (#1);
	\end{tikzpicture}}
\newcommand*\fullcirc[1][1ex]{\tikz\fill (0,0) circle (#1);}

\AtBeginDocument{%
  \providecommand\BibTeX{{%
    \normalfont B\kern-0.5em{\scshape i\kern-0.25em b}\kern-0.8em\TeX}}}

\def\oursys{MVP-Shapley\xspace}

\begin{document}

\title{\oursys: Feature-based Modeling for Evaluating the Most Valuable Player in Basketball}

\author{Haifeng Sun\textsuperscript{*}, Yu Xiong\textsuperscript{\dag}, Runze Wu\textsuperscript{\dag}, Kai Wang\textsuperscript{*}, Lan Zhang\textsuperscript{*}, Changjie Fan\textsuperscript{\dag}, Shaojie Tang\textsuperscript{+}, Xiang-Yang Li\textsuperscript{*}}
\affiliation{%
  \institution{\textsuperscript{*}University of Science and Technology of China; \textsuperscript{\dag}Netease, Fuxi AI Lab; \textsuperscript{+} University at Buffalo }
  \country{\textsuperscript{*}Hefei, China; \textsuperscript{\dag}Hangzhou, China; \textsuperscript{+}Buffalo, New York, USA }
}
\email{{sun1998, kaiwang7}@mail.ustc.edu.cn; {zhanglan, xiangyangli}@ustc.edu.cn}
\email{{xiongyu1, wurunze1, 
fanchangjie}@corp.netease.com; shaojiet@buffalo.edu}








\input{1_Abstract}


\maketitle

\input{2_Introduction}

\input{4_Preliminary}

\input{5_Framework}

\input{6_Experiment}

\input{3_Related_work}

\input{8_Conclusion}


\clearpage
\bibliographystyle{ACM-Reference-Format}
\bibliography{sample-base}

\appendix
\input{9_appendix}

\end{document}

%% file: 1_Abstract.tex
\begin{abstract}

The burgeoning growth of the esports and multiplayer online gaming community has highlighted the critical importance of evaluating the Most Valuable Player (MVP). 
The establishment of an explainable and practical MVP evaluation method is very challenging.
In our study, we specifically focus on play-by-play data, which records related events during the game, such as assists and points.
We aim to address the challenges by introducing a new MVP evaluation framework, denoted as \oursys, which leverages Shapley values. This approach encompasses feature processing, win-loss model training, Shapley value allocation, and MVP ranking determination based on players' contributions. Additionally, we optimize our algorithm to align with expert voting results from the perspective of causality. We also provide a rigorous theoretical analysis with fairness guarantees. Finally, we substantiate the efficacy of our method through validation using the NBA dataset and the Dunk City Dynasty dataset and implemented online deployment in the industry, achieving performance improvements of 1.42$\times$ and 1.44$\times$, respectively.
Our code is available at~\url{https://anonymous.4open.science/r/MVP-Shapley-F18B/} and our data is available at~\url{https://anonymous.4open.science/r/nba-data-1078}.
\end{abstract}

%% file: 2_Introduction.tex
\section{Introduction}
  
The rapid growth of esports and online multiplayer gaming underscores the need for fair Most Valuable Player (MVP) evaluation. MVPs exemplify superior skill, tactics, and teamwork, driving team victories. Beyond prestige, such recognition motivates excellence, boosts team performance, and cultivates competitive ecosystems. A fair, scalable MVP framework is thus vital for sustainable esports development.
In basketball analytics, data falls into two types: (i) \emph{tracking data}, captured via optical or device systems for player/ball trajectories~\cite{gudmundsson2017spatio}; (ii) \emph{play-by-play data}, logging game events like assists. Due to high costs~\cite{decroos2019actions}, we focus on play-by-play data.
\begin{figure}[thbp]
  \centering
  \includegraphics[width=\linewidth]{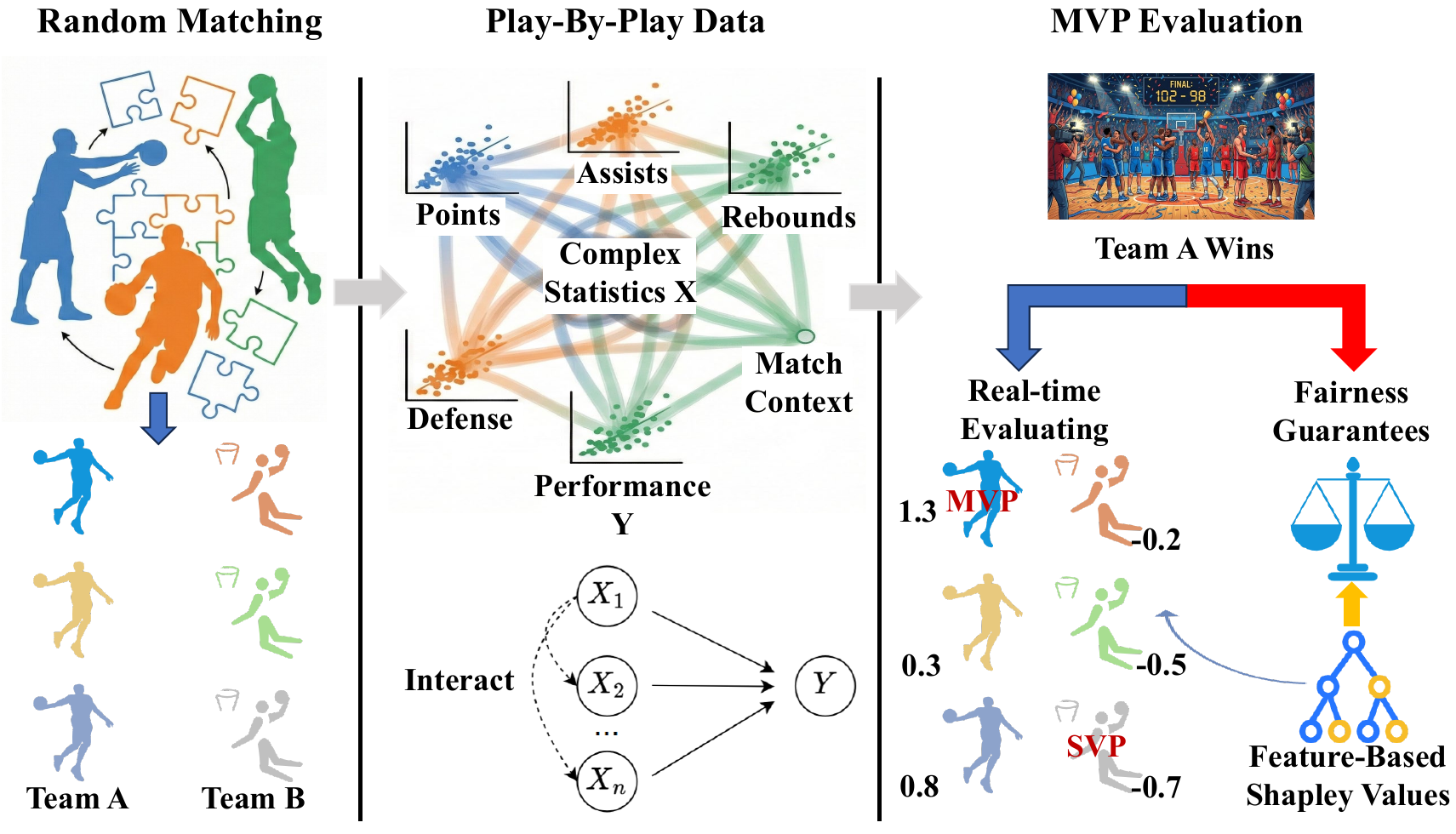}
  \caption{Illustration of the MVP Evaluation of online basketball games.}
  \label{mvp_intro}
\end{figure}
\begin{table*}[htbp]
\setlength{\belowcaptionskip}{0.15cm}
\caption{Comparison with Existing Works.}
\label{Tab:compare}
\resizebox{0.98\linewidth}{!}{
\begin{threeparttable}

\centering
\begin{tabular}{@{}cccccc@{}}
\toprule
\textbf{Method} &\textbf{Scalability$^*$} &\textbf{Explainability} & \textbf{Collinearity Issue} & \textbf{High Costs$^\dagger$} & \textbf{Real-time}\\ \midrule
 Metric Weighting  & \checkmark  & \emptycirc  & \checkmark   &  $\times$ & \checkmark \\
 Voting Selection & \checkmark & \fullcirc & $\times$   & \checkmark & $\times$ \\
 Machine Learning Techniques &  \checkmark & \halfcirc  & \checkmark   & $\times$ & \checkmark\\
 Cooperative Game Theory among Players & $\times$ & \fullcirc  & $\times$  & $\times$ & \checkmark \\

 Ours & \checkmark & \fullcirc  & $\times$  & $\times$ & \checkmark \\ 
\bottomrule
\end{tabular}

    \begin{tablenotes}
        \small
        \item 1. $\fullcirc$, $\checkmark$: Yes; $\emptycirc$, $\times$: No; $\halfcirc$: Partial, which means a significant gap exists between the current state and the desired goal.
        \item 2. $^*$: Applicable to evaluating single-game MVP, regular season MVP, and other games with play-by-play data.
        \item 3. $^\dagger$: The cost includes manpower, material resources, and money.
      \end{tablenotes}
\end{threeparttable}
}
\end{table*}

Existing MVP methods fall into four categories:
\textbf{Metric Weighting.} Methods like PM, APM~\cite{madhavan2016predicting,sill2010improved}, BPM~\cite{kubatko2007starting,grassetti2021extended}, RPM~\cite{engelmann2017possession}, WS or WS48~\cite{cao2012sports}, WARP, and VORP~\cite{sarlis2020sports} assign empirical weights to metrics~\cite{cooper2009selecting,piette2010scoring}. Industry examples include Netease's ``Dunk City Dynasty'' (points, assists, etc.) and Tencent's ``Honor of Kings'' (KDA-based). However, these methods lack interpretability, as weights are arbitrarily set based on experience, potentially leading to high-weighted metrics not necessarily contributing significantly to the win rate. Additionally, there may be collinearity among metrics, making MVP calculation challenging. 
\textbf{Voting Selection.} Experts vote based on stats and observation (e.g., NBA). We assume votes are fair ground truth, but the process is time-consuming and costly.
\textbf{Machine Learning.} Black-box models predict value~\cite{fearnhead2011estimating,page2013effect,metulini2020measuring,sandri2020markov,terner2021modeling}, often ignoring collinearity and interpretability.
\textbf{Cooperative Game Theory.} Shapley values assess contributions via lineup permutations~\cite{metulini2023measuring,kolykhalova2020automated,matthiopoulou2020computational}. However, their approach is limited to assessing the value of players within a single, stable team. Furthermore, this method is not applicable for evaluating the MVP in an online basketball game, where the team formation may be random and unstable.

Our study faces two main challenges, as Figure~\ref{mvp_intro} illustrates: 
1). Online games with random team formations have limited historical combinations of players, making it difficult to apply the leave-one-out technique for players by traditional Shapley methods.
2). Player statistics are complex and diverse, with the presence of confounding variables, which complicates the evaluation process.

To address the limitations of existing MVP evaluations (see Table~\ref{Tab:compare}), which include poor interpretability, sensitivity to collinearity, high computational or human costs, and limited scalability, we propose \oursys, a novel Shapley value-based framework for fair and real-time MVP assessment in team-based competitive games.
\oursys operates in five main stages.
     \textbf{Feature preprocessing}: We construct a paired win/loss representation from play-by-play statistics, doubling the effective sample size while preserving game symmetry.
     \textbf{Win-loss prediction}: A high-performance LightGBM model is trained to predict game outcomes, serving as the utility function for subsequent attribution.
     \textbf{Player contribution computation}: TreeSHAP is employed to compute exact feature-level Shapley values with respect to the predicted win probability; player-level contributions are then obtained by aggregating and differencing home/away marginal effects.
     \textbf{MVP evaluation}: Three ranking strategies are proposed for single-game and multi-game (season/finals) evaluation, balancing consistency and robustness.
     \textbf{Causal refinements}: Confounding variables (especially high-importance but distorting metrics) are identified via feature importance analysis and causal reasoning; fuzzification (discretization into bins) is applied to mitigate their confounding bias and improve alignment with ground-truth MVP votes.

Our main contributions are:

$\bullet$  We introduce \oursys, an interpretable, scalable, and theoretically grounded Shapley-based MVP evaluation framework applicable to both single-game and season-long settings. We provide rigorous guarantees of fairness, consistency, and statistical convergence.

$\bullet$ From a causal perspective, we propose a practical workflow that identifies and mitigates the influence of strong confounders through low-complexity subset search and distribution-aware fuzzification, markedly improving alignment with human voting ground truth.

$\bullet$ We release a comprehensive NBA dataset (2000–present regular season + finals) enriched with basic and advanced statistics.
Additionally, we leverage crowdsourcing to gather MVP rankings for the Dunk City Dynasty dataset through voting, using these rankings as the ground truth to validate our method.

$\bullet$ 
We substantiate the efficacy of \oursys through validation using the NBA dataset and the Dunk City Dynasty dataset, achieving performance improvements of $1.42\times$ and $1.44\times$, respectively. Real-world online deployment has reduced the report rate by approximately $9.64\text{\textperthousand} \pm 1.2\text{\textperthousand}$ and the churn rate by approximately $8.11\text{\textperthousand} \pm 0.9\text{\textperthousand}$ compared to the existing method.

%% file: 4_Preliminary.tex
\section{Preliminaries}
In this work, we focus on the MVP evaluation problem using play-by-play data and Shapley values~\cite{shapley1953value}. In this section, we will introduce the datasets and Shapley values.
\subsection{Dataset}
\subsubsection{NBA Dataset}
We crawled all regular season and playoff play-by-play data of the NBA since 2000 from the basketball reference website (~\url{www.basketball-reference.com/}). 
There are 31,627 games in total, and four tables for each of the two teams in each game: one basic statistics table and one advanced statistics table for each team.
Each game contains 20 basic statistics and 15 advanced statistics of all players on the court. 
Table~\ref{tab:stats_summary} shows the description of basic and advanced statistics.
Our collated dataset is available at~\url{https://anonymous.4open.science/r/nba-data-1078}.

\subsubsection{Dunk City Dynasty Dataset}
Dunk City Dynasty~\footnote{\url{https://www.dunkcitymobile.com/}
} is a 3V3 basketball competitive mobile game developed by Netease and authorized by the NBA official players union. The dataset comprises play-by-play data, including the following 82 statistics such as `BeStrongDisturbedShoot', `NicePass', `BuffUltraSkillSteal', `TwoShots', `Block', `DisturbedLayupNum', and others.
A total of 184,908 games are selected for the Summit match on March 15, 2024.

\subsection{Shapley Values}
The Shapley value (SV)~\cite{shapley1953value} is commonly used in game theory to identify the contributions of players collaborating in a coalition.
Assume there are $n$ players $N=\{1,\cdots, n\}$, $S\subset N$ is a subset of this coalition. 
Given a utility function $v: S \to \mathbb{R}$, the SV of the player $i$ is defined as $\phi_i(v)$, which is the average marginal contribution of $i$ to all possible subsets of $S$:
\begin{equation}
\phi_i(v) = \frac{1}{n}\sum_{S\subset N/\{i\}}{\tbinom{n-1}{|S|}}^{-1}(v(S\cup \{i\})-v(S)).
\label{Shapley}
\end{equation}
Here, $|S|$ represents the size of the subset $S$. The term $\frac{1}{n}{\tbinom{n-1}{|S|}}^{-1}$ represents the probability of the subset $S$ appearing. The term $v(S\cup i)-v(S)$ represents the marginal contribution of player $i$ to the sub-coalition $S$.

%% file: 5_Framework.tex
\section{Methods}
\begin{figure*}[thbp]
  \centering
  \includegraphics[width=\linewidth]{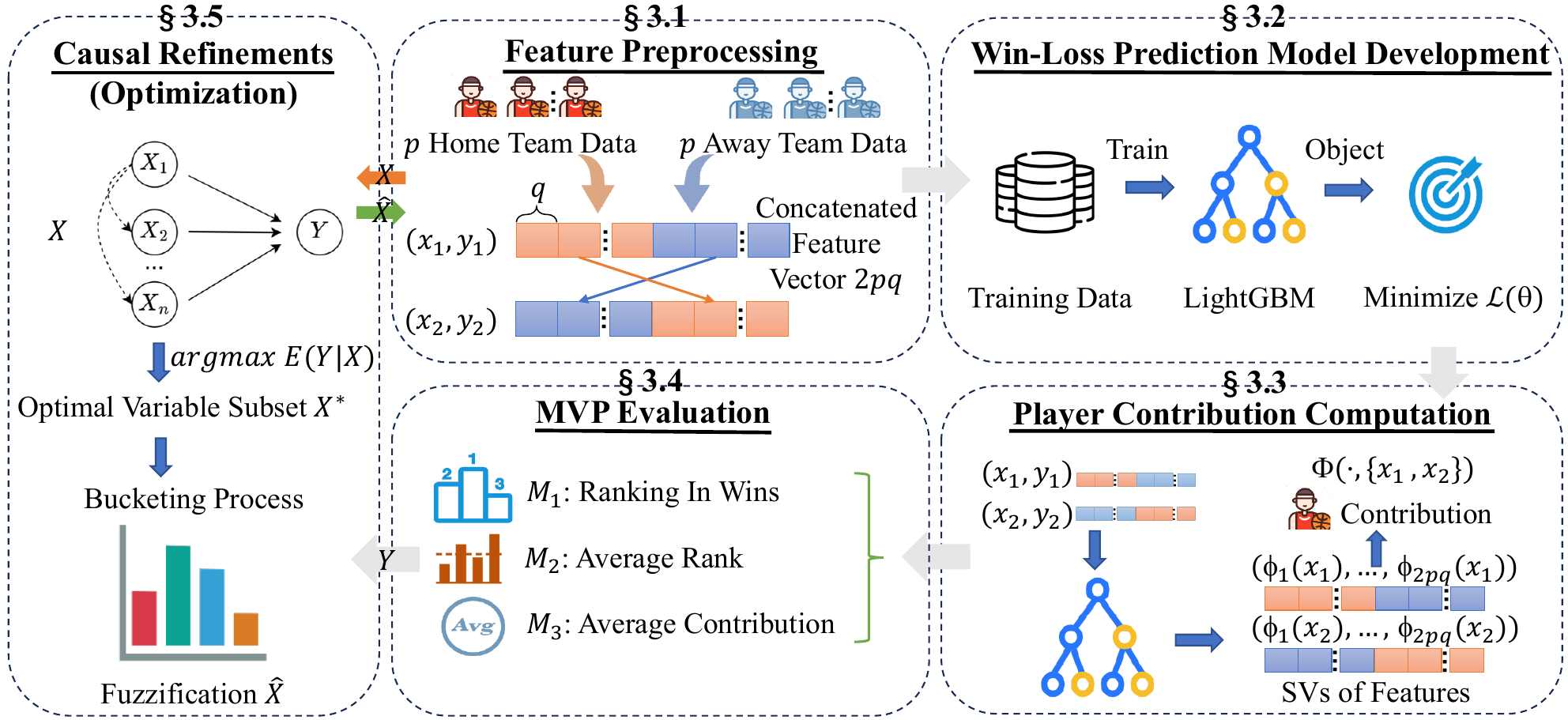}
  \caption{Overview of \oursys.}
  \label{method}
\end{figure*}
\oursys consists of five parts as shown in Figure~\ref{method}: (1) feature preprocessing, (2) win-loss prediction model development, (3) computation of players' contributions, and (4) MVP evaluation. (5) Finally, to improve the algorithm's alignment with empirical voting results, we refine our approach from a causal perspective by incorporating feature combinations and fuzzification.

\subsection{Feature Preprocessing}
\label{Feature_Preprocessing}
Assume the maximum number of players on each team is $p$ with any missing players padded with zeros. Each player's statistical data is represented by a vector of dimension $q$ (including basic and advanced statistics). For each match, the features of all players from both teams are concatenated into a feature vector of dimension $2pq$. The first $pq$ dimensions correspond to the home team's players, followed by the $pq$ dimensions representing the away team's players. The label denotes the home team's game result, with 1 representing a home team win and 0 representing a home team loss. It is crucial to understand that "home team" refers to the team whose features are presented first in the data. For each pair of teams in a game, the outcomes for the winning and losing teams are complementary, enabling the construction of two data points with opposite labels. For instance, if team A defeats team B in a game, let the sets of all players from team A and team B be $\{a_1,\cdots,a_p \}$ and $\{b_1,\cdots,b_p \}$, respectively. The statistical data features of player $a_i$ are recorded as $\{a_{i1},\cdots,a_{iq}\}$, and those of player $b_i$ as $\{b_{i1},\cdots,b_{iq}\}$. Consequently, two data points are constructed for this game:$\{(x_1,y_1),(x_2,y_2)\}$, where $x_1 =(a_{11},\cdots, a_{1q},\cdots, a_{p1},\cdots, a_{pq},\cdots, b_{11},\cdots, b_{1q}, \cdots, b_{p1},\cdots, b_{pq})$, \\$y_1 = 1$, and $x_2 =(b_{11},\cdots, b_{1q},\cdots, b_{p1},\cdots, b_{pq},\cdots, a_{11},\cdots, a_{1q},\\ \cdots, a_{p1},\cdots, a_{pq})$, $y_2 = 0$. The dataset for $m$ games is $\{(x_i,y_i)|i\in \{1,\cdots,2m\}\}$.

\subsection{Win-Loss Prediction Model Development}
After constructing the dataset, we select $n$ data points as the training set. We employ the LightGBM model~\cite{ke2017lightgbm} for predictive analysis. LightGBM is a gradient-boosting framework that utilizes tree-based learning algorithms and is engineered for efficiency and high performance. The model minimizes the following objective function:
\begin{equation}
\mathcal{L}(\theta) = \sum_{i}^n l(y_i, f(x_i)) + \sum_{i}\Omega(f_i) + \sum_{k}\Omega(\gamma_k),
\end{equation}
where $\mathcal{L}(\theta)$ denotes the overall objective function and $\theta$ denotes the model parameters. 
$y_i$ represents the true label, $f(x_i)$ is the predicted score. $f(x_i) = (f(x_i)_{loss}, f(x_i)_{win})\in R^2$, where $f(x_i)_{loss}+ f(x_i)_{win} = 1$, and $f(x_i)_{win}$ represents the probability of winning. $l(y_i, f(x_i))$ denotes the binary log loss function and is defined as:
\begin{equation}
\begin{aligned}
l(y_i, f(x_i)) &= - ( y_i \log(f(x_i)_{win}) + (1 - y_i) \\
&\log(1 - f(x_i)_{win}) ). 
\end{aligned}
\end{equation}
$\Omega(f_i)$ denotes the regularization term for the $i$-th tree, and $\Omega(\gamma_k)$ represents the regularization term for the $k$-th leaf.
The objective function $\mathcal{L}(\theta)$ encapsulates the model's optimization goal, aiming to minimize this function through the learning process.

\subsection{ Player Contribution Computation}

\oursys first computes the Shapley values for the player's features, then uses these values to assess the player's contribution to the match outcome. Inspired by probability-based Shapley Value~\cite{xia2024p}, the winning probability from the win-loss model $f$ is used as the utility function $v$ in the Shapley formula~(\ref{Shapley}): $v(S) = f(x_S)_{win}$.
For a game data $x=(x^1,\cdots,x^{2pq})\in R^{2pq}$, $S$ represents a subset of feature indices, and $x_S$ denotes the feature subset of $x$ corresponds to the indices in $S$.
\oursys calculates the average marginal contribution $\phi_i$ of each feature $x^i$ to the winning score $f(x)_{win}$ across all possible combinations of other features, i.e., the Shapley value, and uses this as the contribution of feature $x^i$ to the winning rate. Let $N=\{1,\cdots,2pq\}$. The calculation of $\phi_i(x)$ is as follows:
\begin{equation}
\phi_i(x) = \frac{1}{2pq}\sum_{S\subset N/\{i\}}{\tbinom{2pq-1}{|S|}}^{-1}(f(x_{S\cup\{i\}})_{win} - f(x_S)_{win}).
\end{equation} 
Next, we will explain how to evaluate a player's contribution in a single game.
We construct two data $x_1$ and $x_2$ as shown in~\ref{Feature_Preprocessing}.
The contribution of the player $a_i$ or $b_i$ to the game outcome can be defined as the sum of all Shapley values of the player's features on the home team minus the sum of all Shapley values of the player's features on the away team. This can be denoted as $\Phi(a_i,\{x_1,x_2\})$ and $\Phi(b_i,\{x_1,x_2\})$, respectively.
\begin{equation}
\begin{aligned}
\Phi(a_i,\{x_1,x_2\}) &= \sum_{j\in\mathcal{H}(a_i)}\phi_j(x_1)-\sum_{j\in\mathcal{A}(a_i)}\phi_j(x_2), \\
\Phi(b_i,\{x_1,x_2\}) &= \sum_{j\in\mathcal{H}(b_i)}\phi_j(x_2)-\sum_{j\in\mathcal{A}(b_i)}\phi_j(x_1).
\end{aligned}
\end{equation} 
Here, $\mathcal{H}(a_i)$ represents the index set of all features of player $a_i$ when he is on the home team.
$\mathcal{A}(a_i)$ represents the index set of all features of player $a_i$ when he is on the away team. Similarly, $\mathcal{H}(b_i)$, $\mathcal{A}(b_i)$ denote the index sets of all features of player $b_i$ when they are on the home and away teams, respectively.



\subsection{MVP Evaluation}
\subsubsection{MVP evaluation of a single game}
For a single game, generally speaking, the MVP is awarded to the player from the winning team. Since team A defeated team B, the player with the highest contribution value calculated from team A is the MVP of this game. 
The calculation is as follows:
\begin{equation}
MVP = \arg\max_{a_i}\Phi(a_i,\{x_1,x_2\}).
\end{equation} 

\subsubsection{MVP evaluation of multiple games}
~\label{MVP_evaluation_methods}
However, the evaluation of the regular season MVP and the finals MVP  is based on the comprehensive performance of players across multiple games.
To address this, \oursys provides three distinct methods for calculating the MVP across multiple games. Let $p_i$ represent a player who has participated in $T$ games, with $\Phi_i(p_i)$ denoting the player's contribution to the $i$-th game. As the previous section discussed, we have established a method for calculating the contribution of a player in a single game.

1) Method 1 (\textbf{$M_1$}): The MVP is determined by the player with the smallest average ranking of their contribution across all winning games they participated in. Let $G_{win}$ represent the set of winning games for player $p_i$. The MVP is calculated as follows:
\begin{equation}
MVP = \arg\min_{a_i} \frac{1}{|G_{win}|}\sum_{i\in G_{win}}rank(\Phi_i(p_i)).
\end{equation} 
Here, $rank(\Phi_i(p_i))$ represents the ranking of player $p_i$’s contribution $\Phi_i(p_i)$ in this game from large to small.

2) Method 2 (\textbf{$M_2$}): The MVP is determined by the player with the smallest average ranking of their contribution across all games they participated in, regardless of the outcome. This method evaluates the player's performance uniformly across all games. The MVP is calculated as follows:
\begin{equation}
MVP = \arg\min_{a_i} \frac{1}{T}\sum_{i\in \{1,\cdots,T\}}rank(\Phi_i(p_i)).
\end{equation} 

3) Method 3 (\textbf{$M_3$}): The MVP is determined by the player with the largest average contribution across all games they participated in. This method focuses on the sum of the Shapley values themselves, rather than their rankings. The MVP is calculated as follows: 
\begin{equation}
MVP = \arg\max_{a_i} \frac{1}{T}\sum_{i\in \{1,\cdots,T\}}\Phi_i(p_i).
\end{equation} 

\subsection{Causal Refinements}

In our analysis, we possess the ground truth of the MVP voting data. These statistics are processed into features to train a win-loss prediction model and compute a utility function. However, it remains unclear which features truly act as causal drivers of the ground-truth outcomes. Let \( Y \) represent the ground truth, and \( X_1, X_2, \ldots, X_n \) denote the variables related to computing the utility function before feature preprocessing. Our goal is to identify the optimal feature subset \( X^* \) that maximizes the conditional expectation of the outcome:
$X^* = \underset{X}{\arg\max} \, \mathbb{E}[Y | X].$
Exhaustive search over all $2^n$ subsets is computationally prohibitive. To reduce complexity, we first apply feature importance ranking to group low-importance features together. 
For instance, Figure~\ref{feature_v0} shows feature importance on the NBA win-loss model. 
The top two most influential variables (`+,-' and `DRtg') are treated individually, and all others are grouped, yielding only 8 candidate combinations.


Statistical variables largely capture players' in-game behavioral performance and contribute to the final outcome. However, some variables—known as confounders—not only affect the outcome directly but also influence other variables, thereby distorting their apparent importance.To reduce this confounding effect and allow other features to reflect their true contribution more accurately, we apply fuzzification by discretizing continuous variables into bins.
For a variable $X_i$ (e.g., `DRtg'), we partition its values into $t$ intervals defined by boundary points $B = \{bin_1, bin_2, \dots, bin_t\}$. The discretized version $\hat{X}_i$ is then defined as:
\begin{equation}
 \hat{X_i} = \begin{cases}
1 & \text{if } X_i \leq bin_1 \\
j & \text{if } bin_{j-1} < X_i \leq bin_j, 
\end{cases}
\end{equation} 
The resulting $\hat{X}_i$ (the fuzzified DRtg) is used in place of the original value to lessen its excessive influence on MVP determination. The bin boundaries $B$ are chosen according to the empirical distribution of DRtg, ensuring effective control of its confounding effect.
When $|B| = 1$, the variable is effectively removed from consideration.
Figure~\ref{feature_v10} illustrates the feature importance distribution after fuzzification (with $|B|=3$), demonstrating a noticeably more balanced contribution across variables.


\subsection{Theoretical Analysis}
In this section, we provide a rigorous theoretical analysis of the \oursys framework. We first 
establish the fairness properties of our MVP evaluation framework, including fairness guarantees (Theorem~\ref{thm:Fairness}) and consistency analysis (Theorem~\ref{thm:m1}).

We then highlight the uniqueness of our approach among attribution methods.

\begin{theorem}[Uniqueness Theorem]
The \oursys player contribution measure $\Phi$ is the unique attribution method at the feature level satisfying: {Efficiency} (Theorem~\ref{thm:efficiency_single}); {Symmetry} (Theorem~\ref{thm:symmetry}); {Null Player} (Theorem~\ref{thm:null}); {Additivity} (Theorem~\ref{thm:additivity}).
\end{theorem}

Next, we analyze the convergence properties of our approach as the number of games increases.

\begin{theorem}[Law of Large Numbers for MVP Ranking]
Let $\mu_i = \mathbb{E}[\Phi(p_i)]$ be the expected contribution of player $p_i$. For method $M_3$, as the number of games $T \rightarrow \infty$:
$\bar{\Phi}_T(p_i) = \frac{1}{T}\sum_{t=1}^{T}\Phi_t(p_i) \xrightarrow{a.s.} \mu_i,$
where $\xrightarrow{a.s.}$ denotes almost sure convergence.
\end{theorem}
In addition, we also analyzed the concentration bound (Theorem~\ref{thm:concentration}) and sample complexity for MVP identification (Corollary~\ref{cor:sample}).
\begin{theorem}[Concentration Bound]
For any player $p_i$ with bounded contribution $|\Phi_t(p_i)| \leq M$, with probability at least $1-\delta$:
$\left|\bar{\Phi}_T(p_i) - \mu_i\right| \leq M\sqrt{\frac{2\ln(2/\delta)}{T}}.$
\end{theorem}

\begin{corollary}[Sample Complexity for MVP Identification]
To distinguish between two players with expected contribution difference $\Delta = |\mu_i - \mu_j|$ with probability at least $1-\delta$, the required number of games is:
$T \geq \frac{8M^2\ln(4/\delta)}{\Delta^2}.$
\end{corollary}

Finally, we provide a theoretical justification for the fuzzification approach to handle confounding variables.


\begin{theorem}[Bias Reduction through Fuzzification]
Let $X_c$ be a confounding variable with range $[a,b]$. The fuzzified variable $\hat{X}_c$ with $t$ bins reduces the mutual information between $\hat{X}_c$ and other variables:
$I(\hat{X}_c; X_i) \leq I(X_c; X_i),$
where $I(\cdot;\cdot)$ denotes mutual information.
\end{theorem}

These theoretical analysis provides a solid mathematical foundation for \oursys, demonstrating that our approach offers fair, consistent, and theoretically justified MVP evaluations based on the well-established Shapley value framework from cooperative game theory.
For more details and proof, please refer to APPENDIX~\ref{Proof}.

%% file: 6_Experiment.tex
\section{Experiments}
\subsection{Experimental Setup}
During the training of the win-loss model, we split the dataset into a training set and a test set with a 9:1 ratio. The models trained on the NBA and Dunk City Dynasty datasets achieved accuracies of 0.99 and 0.97, respectively.
Here, we employ TreeSHAP~\cite{lundberg2020local} to compute the Shapley values efficiently. 
For the NBA dataset, we divide the variables into five groups according to their importance: `+/-', `ORtg', `DRtg', `BPM', and the remaining variables. 
\oursys has three versions with different utility functions. Different versions train the win-loss models by constructing different features. 
The first version (\textbf{Ours\_$\mathbf{V_1}$}) uses all features. 
The second version (\textbf{Ours\_$\mathbf{V_2}$}) uses the remaining features after removing `+,-' and `DRtg'. 
The third version (\textbf{Ours\_$\mathbf{V_3}$}) uses the
features after fuzzification. For `+/-', we set $|B|=3$ and for `DRtg', $|B|=8$.
Here, the second version focuses solely on combinatorial optimization, while the third version incorporates fuzzy optimization. For the Dunk City Dynasty dataset, we use the optimized features for analysis in Section~\ref{l33_results}. 
See the appendix for more dataset analysis.
\subsubsection{Baselines}
1) \textbf{Ground Truth (GT):}
\textbf{NBA Dataset:} The NBA regular season MVP rankings and Finals MVP awards are used as the ground truth, with detailed information available on the website: ~\url{https://www.basketball-reference.com/awards}. 
\textbf{Dunk City Dynasty Dataset:} For the Dunk City Dynasty dataset, crowdsourcing is used for MVP ranking voting on the \textit{NetEase Youling Crowdsourcing Platform} (~\url{https://zb.163.com/mark/task}). Annotators, who are professional basketball enthusiasts, carefully review game videos and rank the winning players (A total of $3!=6$ rankings), considering factors beyond just points, assists, defense, rebounds, steals, and turnovers, such as overall contributions to the game. Annotators should objectively choose the best one from 6 ranking options. 
A total of 500 game videos are included in our crowdsourcing effort.
The platform employs truth inference algorithms~\cite{dawid1979maximum, weng2024humandata} to establish crowdsourced confidence.

2) \textbf{GSv~\cite{metulini2023measuring}:}
According to the team's historical lineup data, GSv fits different lineup win rate prediction models and uses the generalized Shapley value to calculate each team's MVP and best lineup.

3) \textbf{API~\cite{sarlis2020sports}:}
This is a metric weighting method; each advanced efficiency metric is normalized and weighted to obtain the API. 

\subsubsection{Evaluation Metrics}


1) \textbf{Average Rank Difference (ARD):}
The average rank difference quantifies the average discrepancy between the predicted MVP rank and the actual rank of the corresponding player. Mathematically, ARD is calculated as follows:
$ARD = \frac{1}{N} \sum_{i=1}^{N} |rank_{\text{predicted MVP}_i} - rank_{\text{ground truth}_i}|.$
$N$ is the total number of players on the MVP voting list.
$rank_{\text{predicted MVP}_i}$ is the rank of the predicted MVP for the i-th player on the MVP voting list, 
$rank_{\text{ground truth}_i}$ is the actual voting rank of the i-th player.

2) \textbf{Spearman's Rank Correlation Coefficient~\cite{sedgwick2014spearman} (SRCC):}
Spearman's Rank Correlation Coefficient is a measure of the strength and direction of association between two ranked variables. 
The formula for Spearman's Rank Correlation Coefficient is given by:
$ SRCC = 1 - \frac{6 \sum d_i^2}{n(n^2 - 1)}. $
Here, $d_i$ represents the difference between the ranks of corresponding players, and $n$ is the number of players. Spearman's Rank Correlation Coefficient ranges from -1 to 1, where 1 indicates a perfect positive monotonic relationship, and -1 indicates a perfect negative monotonic relationship.

3) \textbf{ Recall (R):}
The recall here is defined as the proportion of players ranked in the top K of the MVP voting appearing in the predicted top K by one MVP evaluation method. Let $G_K$ represent the set of top K players in MVP voting, and $M_K$ represent the set of top K players predicted by the method. 
This can be expressed as:
$R = \frac{|G_K\bigcap M_K|}{K}.$


4) \textbf{ Accuracy (ACC):}
Assuming there are \( n \) matches, each with an MVP voting result. Let \( m \) denote the number of matches for which one method's evaluation for MVP matches the voting result. The accuracy can be expressed as: 
$\text{ACC} = \frac{m}{n}.$


\subsection{Results}

\subsubsection{NBA Regular Season MVP Results}

We analyze the outcomes of the regular season MVP experiments over the past three years. 
$\mathbf{M_1}$, $\mathbf{M_2}$, and $\mathbf{M_3}$ correspond to the MVP evaluation method in Section~\ref{MVP_evaluation_methods}. We test three evaluation metrics of these methods on three different versions of the win-loss model.
Table~\ref{table_metrics_topall} presents the performance comparison across all ranked players in the 2022--2024 NBA regular season MVP voting results, while Table~\ref{table_metrics_top3} focuses on the top-three ranked players.
Our best variant, \textbf{Ours\_V2} ($M_3$), achieves substantial improvements over the baselines (GSV and API) and over the previous version \textbf{Ours\_V1} ($M_3$). On the full ranking (Table~\ref{table_metrics_topall}), Ours\_V2 ($M_3$) reduces ARD by up to \textbf{21.35--35.80$\times$} compared to GSV and by \textbf{4.80--8.29$\times$} compared to API. It also yields a \textbf{0.42--1.91$\times$} improvement in
SRCC, and a \textbf{0.60--2.05$\times$} relative gain in R. For top-3 ranking tasks (Table~\ref{table_metrics_top3}), Ours\_V2 ($M_2$ and $M_3$) consistently attains near-perfect or perfect SRCC (1.00) and R (0.67--1.00) in multiple seasons, significantly outperforming GSV (SRCC $\leq$ 0.50, R = 0.00) and API (R $\leq$ 0.33), while reducing ARD by large margins (e.g., 0.00--2.33 vs.\ 169.00--297.00 for GSV).
These results demonstrate that incorporating causal inference (Ours\_V2) markedly enhances ranking accuracy and stability over both baselines and the prior version, with further refinements via fuzzification (Ours\_V3) providing complementary gains in certain top-3 scenarios.

\begin{table}[htbp]
\newcommand{\tabincell}[2]{\begin{tabular}{@{}#1@{}}#2\end{tabular}}
\renewcommand\arraystretch{1.1}
\centering
\caption{The top-all-ranked players during the
2022-2024 NBA regular season MVP results.}
\resizebox{0.5\textwidth}{!}{
\begin{tabular}{ cccccccccc } 
\hline
\multirow{2}*{\bf{Method }}&\multicolumn{3}{c}{2023/2024} &\multicolumn{3}{c}{2022/2023} &\multicolumn{3}{c}{2021/2022} \\
& \bf{ARD $\downarrow$} & \bf{SRCC $\uparrow$} & \bf{R $\uparrow$} 
& \bf{ARD $\downarrow$} & \bf{SRCC $\uparrow$} & \bf{R $\uparrow$} 
& \bf{ARD $\downarrow$} & \bf{SRCC $\uparrow$} & \bf{R $\uparrow$}   \\
\hline
$M_1$ (Ours\_$V_1$) &  9.00 & \bf{0.93} & 0.56 & 25.00 & 0.41 & 0.31 & 23.42 & 0.39 & 0.33 \\ 
$M_2$ (Ours\_$V_1$) & 15.72 & 0.82 & 0.33 & 44.42 & 0.54 & 0.31 & 39.17 & 0.62 & 0.58   \\ 
$M_3$ (Ours\_$V_1$) & 14.00 & 0.72 & 0.33 & 44.54 & 0.58 & 0.31 & 34.50 & 0.35 &  0.58  \\ 
$M_1$ (Ours\_$V_2$) & 17.89 & 0.68 & \bf{0.67} & 11.85 & 0.57 & 0.54 & 19.92 & 0.32 & 0.50   \\ 
$M_2$ (Ours\_$V_2$) & 12.56 & 0.76 & 0.56 & 8.54 & 0.67 & 0.62 & 23.08 & 0.42 &  0.50  \\ 
$M_3$ (Ours\_$V_2$) & \bf{6.22} & 0.85 & \bf{0.67} & \bf{5.92} & 0.59 & \bf{0.77} & \bf{9.25} & 0.43 &  \bf{0.67}  \\ 
$M_1$ (Ours\_$V_3$) & 26.22 & 0.58 & 0.22 & 33.92 & 0.68 & 0.46 & 39.75 & 0.56 &  0.33  \\ 
$M_2$ (Ours\_$V_3$) & 23.89 & 0.72 & 0.33 & 47.77 & \bf{0.78} & 0.23 & 44.17 & 0.58 &  0.42  \\ 
$M_3$ (Ours\_$V_3$) & 20.67 & 0.77 & 0.33 & 48.54 & 0.68 & 0.31 & 36.17 & \bf{0.74} &  0.42  \\ 
GSV  & 228.89 & -0.60  & 0.00 & 212.84 & 0.32 & 0.00 & 206.75 & -0.13 &  0.00\\ 
API  & 36.64 & 0.32 & 0.22 & 55.00 & 0.45 & 0.46 & 53.67 & 0.52 & 0.42 \\ 

\hline
\end{tabular}
}
\label{table_metrics_topall}
\end{table}

\begin{table}[htbp]
\newcommand{\tabincell}[2]{\begin{tabular}{@{}#1@{}}#2\end{tabular}}
\renewcommand\arraystretch{1.1}
\centering
\caption{The top-three-ranked players during the
2022-2024 NBA regular season MVP results.}
\resizebox{0.5\textwidth}{!}{
\begin{tabular}{ cccccccccc } 
\hline
\multirow{2}*{\bf{Method }}&\multicolumn{3}{c}{2023/2024} &\multicolumn{3}{c}{2022/2023} &\multicolumn{3}{c}{2021/2022} \\
& \bf{ARD $\downarrow$} & \bf{SRCC $\uparrow$} & \bf{R $\uparrow$} 
& \bf{ARD $\downarrow$} & \bf{SRCC $\uparrow$} & \bf{R $\uparrow$} 
& \bf{ARD $\downarrow$} & \bf{SRCC $\uparrow$} & \bf{R $\uparrow$}   \\
\hline
$M_1$ (Ours\_$V_1$)  & \bf{0.33} & \bf{1.00} & \bf{0.67} & 9.67 & \bf{0.50} & 0.33 & 6.67 & -0.50 & 0.33 \\ 
$M_2$ (Ours\_$V_1$)  & 3.00 & \bf{1.00} &  \bf{0.67}& 1.67 & \bf{0.50} & 0.67 & 4.00 & 0.50 & 0.33 \\ 
$M_3$ (Ours\_$V_1$)  & 4.67 & \bf{1.00} & \bf{0.67} & \bf{0.67} & \bf{0.50} & \bf{1.00} & 6.67 & -0.50 & 0.00 \\ 
$M_1$ (Ours\_$V_2$)  & 2.00 & \bf{1.00} & \bf{0.67} & 2.67 & \bf{0.50} & 0.67 & 1.33 & -0.50 & \bf{1.00} \\ 
$M_2$ (Ours\_$V_2$)  & 0.67 & \bf{1.00} & \bf{0.67} & 2.33 & \bf{0.50} & 0.67 & \bf{0.00} & \bf{1.00} & \bf{1.00} \\ 
$M_3$ (Ours\_$V_2$)  & 0.67 & 0.50 & \bf{0.67} & 2.0  & \bf{0.50} & 0.67 & 0.67 & 0.50 & \bf{1.00} \\ 
$M_1$ (Ours\_$V_3$)  &  10.33& \bf{1.00} & 0.33 & 4.67 & -0.50 & 0.33 & 2.00 & -0.50 & 0.67  \\ 
$M_2$ (Ours\_$V_3$)  & 5.33 & \bf{1.00} & 0.33 & \bf{0.67} & \bf{0.50} & \bf{1.00} & 1.00 & -0.50 &  0.67\\ 
$M_3$ (Ours\_$V_3$)  & 4.00 & \bf{1.00} & 0.33 & 1.67 & -1.00 & 0.67 & 1.00 & \bf{1.00} &  0.67\\ 
GSV  & 297.0 & -0.50 & 0.00 & 169.00 & -0.50 & 0.00 & 198.00 & 0.50 & 0.00 \\ 
API  & 22.33 & -0.50 & 0.33 & 5.33 & -0.50 & 0.33 & 2.67 & 0.50 & 0.33 \\ 

\hline
\end{tabular}
}
\label{table_metrics_top3}
\end{table}

Table~\ref{table_regular} shows the MVP selection results of various methods for the 2023/2024 season. 
$ACRW$ represents the average player's contribution ranking across all winning games.
$ACRA$ represents the average player's contribution ranking across all games.
$AC$ represents the average player's contribution across all games.
Our methods all successfully select the MVP of the season except $M_1$ (Ours\_$V_3$).

\subsubsection{NBA Finals MVP Results}
We compute the NBA's Finals Most Valuable Players (FMVPs) over the past decade, presenting the detailed results in Table~\ref{table_final}. We conduct tests using three versions of the win-loss model across three MVP evaluation methods. Subsequently, we document the corresponding FMVP rankings for each method. A smaller ranking indicates closer proximity to the ground truth, with a ranking of 1 signifying that the MVP evaluated by this method aligns with the FMVP of the year.
Our experimental findings demonstrate that the first version of the win-loss model yields the least favorable outcomes, whereas the third version exhibits the highest consistency with the ground truth. Interestingly, this outcome slightly deviates from regular season results, suggesting that there is a difference in how defensive efficiency is considered between regular season and playoff voting.

\begin{table*}[htbp]
\centering
\renewcommand{\arraystretch}{1.1}
\begin{minipage}{0.48\textwidth}
\centering
\caption{The top-three-ranked players during the 2023/2024 NBA regular season MVP results.}
\resizebox{\textwidth}{!}{
\begin{tabular}{cccccc}
\hline
\multirow{2}*{\bf{Rank}} & \multicolumn{1}{c}{GT} & \multicolumn{2}{c}{GSv} & \multicolumn{2}{c}{API} \\
& \bf{Player} & \bf{Player} & \bf{Contribution $\uparrow$} & \bf{Player} & \bf{Contribution $\uparrow$} \\
\hline
1 & Nikola Jokić & Jaylen Martin & 0.995 & Luka Dončić & 0.335 \\ 
2 & Shai Gilgeous-Alexander & Wenyen Gabriel & 0.989 & Sam Hauser & 0.332 \\ 
3 & Luka Dončić & Robert Williams & 0.980 & Victor Wembanyama & 0.331 \\ 
\hline
\end{tabular}
}
\resizebox{\textwidth}{!}{
\begin{tabular}{ccccccc} 
\hline
\multirow{2}*{\bf{Rank}} & \multicolumn{2}{c}{$M_1$ (Ours\_$V_1$)} & \multicolumn{2}{c}{$M_2$ (Ours\_$V_1$)} & \multicolumn{2}{c}{$M_3$ (Ours\_$V_1$)} \\
& \bf{Player} & \bf{ACRW $\downarrow$} & \bf{Player} & \bf{ACRA $\downarrow$} & \bf{Player} & \bf{AC $\uparrow$} \\
\hline
1 & Nikola Jokić & 2.67 & Nikola Jokić & 5.52 & Nikola Jokić & 2.78 \\ 
2 & Shai Gilgeous-Alexander & 3.02 & Kentavious Caldwell-Pope & 5.95 & Shai Gilgeous-Alexander & 2.31 \\ 
3 & Paul George & 3.16 & Shai Gilgeous-Alexander & 6.01 & Kentavious Caldwell-Pope & 2.11 \\ 
\hline
\end{tabular}
}
\resizebox{\textwidth}{!}{
\begin{tabular}{ccccccc} 
\hline
\multirow{2}*{\bf{Rank}} & \multicolumn{2}{c}{$M_1$ (Ours\_$V_2$)} & \multicolumn{2}{c}{$M_2$ (Ours\_$V_2$)} & \multicolumn{2}{c}{$M_3$ (Ours\_$V_2$)} \\
& \bf{Player} & \bf{ACRW $\downarrow$} & \bf{Player} & \bf{ACRA $\downarrow$} & \bf{Player} & \bf{AC $\uparrow$} \\
\hline
1 & Nikola Jokić & 2.38 & Nikola Jokić & 3.23 & Nikola Jokić & 7.39 \\ 
2 & Giannis Antetokounmpo & 2.97 & Giannis Antetokounmpo & 3.47 & Giannis Antetokounmpo & 6.14 \\ 
3 & Shai Gilgeous-Alexander & 3.49 & Shai Gilgeous-Alexander & 4.17 & Luka Dončić & 5.76 \\ 
\hline
\end{tabular}
}
\resizebox{\textwidth}{!}{
\begin{tabular}{ccccccc} 
\hline
\multirow{2}*{\bf{Rank}} & \multicolumn{2}{c}{$M_1$ (Ours\_$V_3$)} & \multicolumn{2}{c}{$M_2$ (Ours\_$V_3$)} & \multicolumn{2}{c}{$M_3$ (Ours\_$V_3$)} \\
& \bf{Player} & \bf{ACRW $\downarrow$} & \bf{Player} & \bf{ACRA $\downarrow$} & \bf{Player} & \bf{AC $\uparrow$} \\
\hline
1 & Victor Wembanyama & 2.21 & Nikola Jokić & 4.85 & Nikola Jokić & 1.31 \\ 
2 & Nikola Jokić & 2.71 & Victor Wembanyama & 5.23 & Victor Wembanyama & 0.97 \\ 
3 & Daniel Gafford & 2.93 & Rudy Gobert & 5.73 & Rudy Gobert & 0.96 \\ 
\hline
\end{tabular}
}
\label{table_regular}
\end{minipage}
\hfill
\begin{minipage}{0.5\textwidth}
\centering
\caption{NBA finals MVP results.}
\resizebox{\textwidth}{!}{
\begin{tabular}{m{1.2cm}<{\centering}m{1.1cm}<{\centering}m{1.1cm}<{\centering}m{1.1cm}<{\centering}m{1.1cm}<{\centering}m{1.1cm}<{\centering}m{1.1cm}<{\centering}m{1.1cm}<{\centering}m{1.1cm}<{\centering}m{1.1cm}<{\centering}m{1.1cm}<{\centering}} 
\hline
\bf{Method (Rank $\downarrow$)} & 2024 & 2023 & 2022 & 2021 & 2020 & 2019 & 2018 & 2017 & 2016 & 2015\\
\bf{FMVP (GT)} & Jaylen Brown & Nikola Jokić & Stephen Curry & Giannis Antetokounmpo & LeBron James & Kawhi Leonard & Kevin Durant & Kevin Durant & LeBron James & Andre Iguodala \\
\hline
$M_1$ (Ours\_$V_1$) & 6 & 2 & 2 & 4 & 5 & 3 & 1 & 4 & 1 & 2\\ 
$M_2$ (Ours\_$V_1$) & 6 & 3 & 3 & 1 & 4 & 4 & 1 & 4 & 3 & 1\\ 
$M_3$ (Ours\_$V_1$) & 1 & 3 & 3 & 2 & 2 & 4 & 1 & 3 & 3 & 1\\ 
$M_1$ (Ours\_$V_2$) & 5 & 1 & 3 & 1 & 1 & 1 & 1 & 1 & 1 & 1\\ 
$M_2$ (Ours\_$V_2$) & 5 & 1 & 1 & 1 & 1 & 1 & 1 & 1 & 1 & 1\\ 
$M_3$ (Ours\_$V_3$) & 5 & 1 & 1 & 1 & 1 & 1 & 1 & 1 & 1 & 1\\ 
$M_1$ (Ours\_$V_3$) & 2 & 1 & 2 & 1 & 1 & 2 & 1 & 2 & 1 & 1\\ 
$M_2$ (Ours\_$V_3$) & 2 & 1 & 1 & 1 & 1 & 1 & 1 & 2 & 1 & 1\\ 
$M_3$ (Ours\_$V_3$) & 2 & 1 & 1 & 1 & 1 & 1 & 1 & 1 & 1 & 3\\ 
\hline
\end{tabular}
}
\label{table_final}
\end{minipage}
\end{table*}

\subsubsection{Dunk City Dynasty MVP Results}
\label{l33_results}

\paragraph{MVP results with crowdsourced confidence.}

\begin{figure}[htbp]
    \centering
    \subfigure[ARD]{
        \label{compare_ARD} 
        \includegraphics[width=0.47\linewidth]{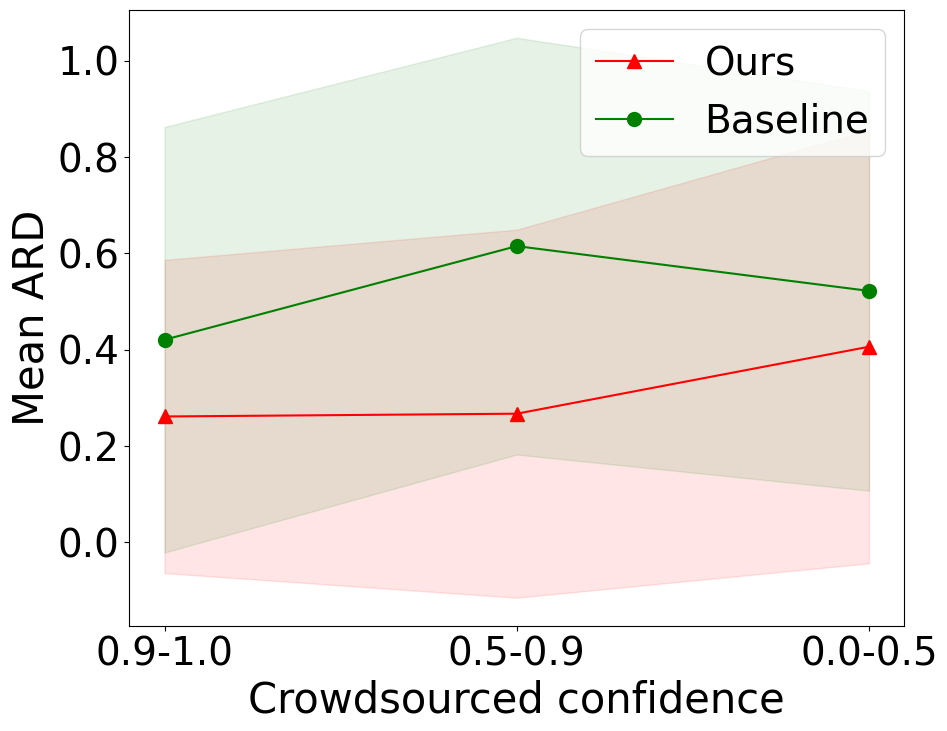}
    }
     \subfigure[SRCC]{
        \label{compare_SRCC} 
        \includegraphics[width=0.47\linewidth]{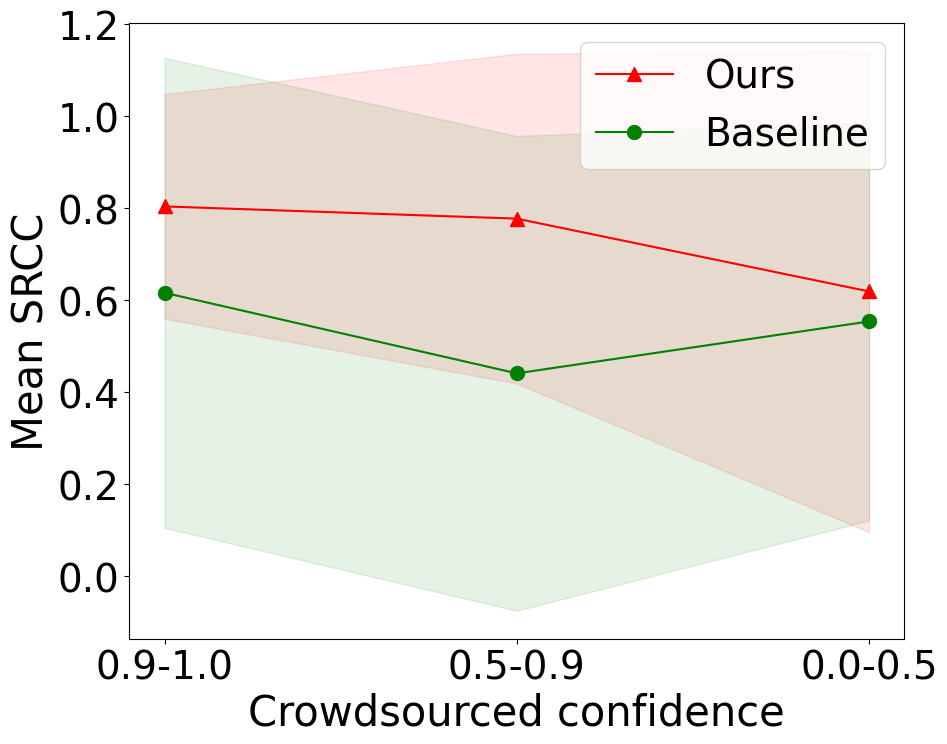}
    }

    \caption{Mean-variance plot of the evaluation metrics to crowdsourced confidence.}  
    \label{l33_confidence_compare} 
\end{figure}

We source a total of 500 videos through crowdsourcing and subsequently consider the results with crowdsourced confidence exceeding 0.9 as the golden dataset. We divide the golden dataset into the dataset required for feature optimization and the test set in a 1:1 ratio. 
Traditional baselines (GSv) are no longer applicable in the Dunk City Dynasty due to the random grouping of players. Dunk City Dynasty currently utilizes an API-based metric-weighted approach for online MVP calculation, which serves as our baseline. 
Specifically, we utilize the `MvpPoint' from the settlement data of each game as the baseline result.
\begin{figure}[htbp]
    \centering
    \subfigure[ARD]{
        \label{l33_baseline} 
        \includegraphics[width=0.47\linewidth]{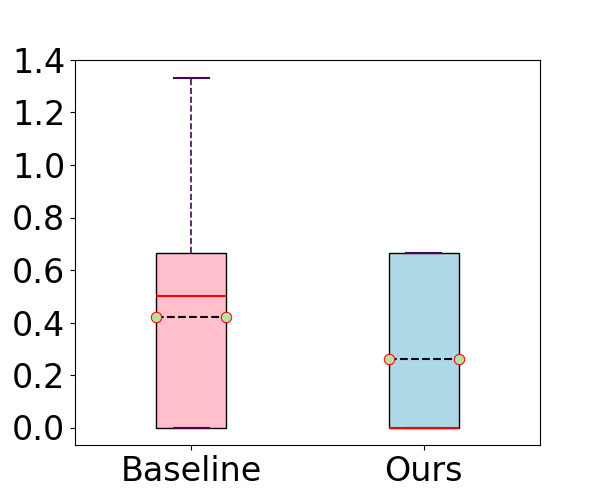}
    }
     \subfigure[SRCC]{
        \label{l33_ours} 
        \includegraphics[width=0.47\linewidth]{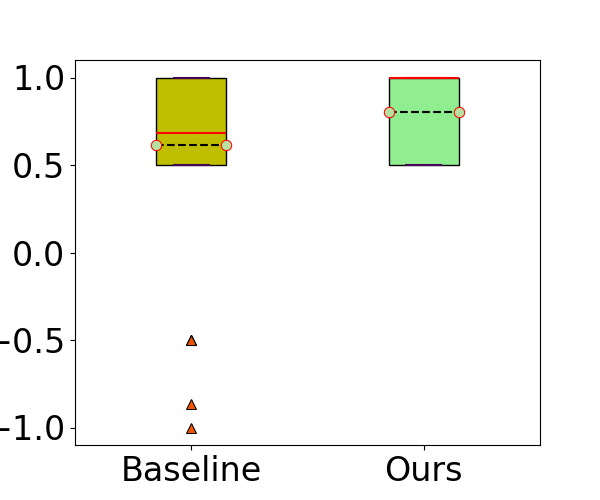}
    }

    \caption{The dashed horizontal line represents the mean, the red horizontal line represents the median, and the orange triangles are outliers.}  
    \label{l33_boxplot} 
\end{figure}
\begin{table}[htbp]
\newcommand{\tabincell}[2]{\begin{tabular}{@{}#1@{}}#2\end{tabular}}
\renewcommand\arraystretch{1.1}
\centering
\caption{The Dunk City Dynasty MVP results.}
\resizebox{0.7\linewidth}{!}{
\begin{tabular}{cccc}
  \hline
  Method & ARD $\downarrow$ & SRCC $\uparrow$ & ACC $\uparrow$\\
  \hline
  Baseline &  0.42 $\pm$ 0.44 & 0.62 $\pm$ 0.51 & 0.54\\
  Ours &  \bf{0.26 $\pm$ 0.33} & \bf{0.80 $\pm$ 0.24} & \bf{0.83}\\
  \hline
\end{tabular}
}
\label{table_l33}
\end{table}
The results of various metrics for Dunk City Dynasty MVP evaluation are presented in Table~\ref{table_l33} and Figure~\ref{l33_boxplot}. Our method significantly outperforms the baseline on the Dunk City Dynasty MVP task, achieving a 38.1\% reduction in ARD, a 29.0\% improvement in SRCC, and a 53.7\% relative gain in ACC.
Furthermore, we explore the influence of crowdsourced confidence on the results. We divide the confidence intervals into 0.0-0.5, 0.5-0.9, 0.9-1.0, and the ratio of the number of games within these three confidence intervals is 1:1:1.
As depicted in Figure~\ref{l33_confidence_compare} and Table~\ref{table_compare_acc}, a decrease in crowdsourced confidence results in a decline in the effectiveness of the MVP evaluation using our method. 
Furthermore, the performance of the Baseline in the confidence interval of 0.0-0.5 is better than that in 0.5-0.9, indicating poor interpretability.

\begin{table}[htbp]
\newcommand{\tabincell}[2]{\begin{tabular}{@{}#1@{}}#2\end{tabular}}
\renewcommand\arraystretch{1.1}
\centering
\caption{Results of accuracy changes with crowdsourced confidence interval variations.}
\resizebox{0.7\linewidth}{!}{
\begin{tabular}{cccc}
  \hline
  Confidence Interval & 0.9-1.0  & 0.5-0.9  & 0.0-0.5 \\
  \hline
  Baseline (ACC $\uparrow$) &  \bf{0.54} & 0.49 & 0.43\\
  Ours (ACC $\uparrow$) &  \bf{0.83} & 0.82 & 0.76\\
  \hline
\end{tabular}
}
\label{table_compare_acc}
\end{table}

\paragraph{MVP results with GOSDT tree~\cite{kang2023learning}.}

However, the golden dataset only accounts for one-fifth of all crowdsourcing results. To fully utilize the voting results from all game videos, we employ the state-of-the-art voting algorithm, the GOSDT tree, to select the MVP for each match. Consequently, every match obtained an MVP ground truth. Ultimately, the obtained ground truth from this algorithm is used to validate the superiority of our approach further. As the voting algorithm selects only one MVP instead of a ranking, we evaluate using two metrics: $ARD$ and $ACC$. The experimental results in Table~\ref{table_l33_GOSDT} demonstrate the superiority of our method, achieving a 40.7\% relative improvement in ARD and a 44.0\% relative gain in ACC.
\begin{table}[htbp]
\newcommand{\tabincell}[2]{\begin{tabular}{@{}#1@{}}#2\end{tabular}}
\renewcommand\arraystretch{1.1}
\centering
\caption{The Dunk City Dynasty MVP results with GOSDT.}
\begin{tabular}{ccc}
  \hline
  Method & ARD $\downarrow$ & ACC $\uparrow$\\
  \hline
  Baseline &  0.59 $\pm$ 0.65  & 0.50\\
  Ours &  \bf{0.35 $\pm$ 0.59} & \bf{0.72}\\
  \hline
\end{tabular}
\label{table_l33_GOSDT}
\vspace{-0.2in}
\end{table}

\subsubsection{Ablation Study}
We evaluate three method variants across all experiments:

\begin{table}[h]
\centering
\caption{Method Variants and Components}
\begin{tabular}{|l|l|}
\hline
\textbf{Version}       & \textbf{Components Included}              \\ \hline
\textbf{Ours\_V1}   & Base model (no optimizations)    \\ \hline
\textbf{Ours\_V2}   & + Causal inference only          \\ \hline
\textbf{Ours\_V3}   & + Causal + Fuzzification         \\ \hline
\end{tabular}
\end{table}



    
The comprehensive experimental results are presented across quantitative comparisons and qualitative analysis. Specifically, Table~\ref{table_regular}--Table~\ref{table_metrics_top3} demonstrate performance improvements at each stage, while Figure~\ref{feature_nba}--Figure~\ref{feature_l33} visualize the feature attribution patterns.
Key findings from the ablation study are as follows. First, using causal inference alone (Ours\_V2) yields a 30.9\% relative improvement in ARD compared to Ours\_V1 in Table~\ref{table_metrics_topall}, improved ranking consistency in SRCC compared to Ours\_V1 in Table~\ref{table_metrics_top3}, and a 19.6\% relative improvement in R compared to Ours\_V1 in Table~\ref{table_metrics_topall}. Second, adding fuzzification (Ours\_V3) further stabilizes the contributions of defensive metrics (as shown in Figure~\ref{feature_nba}) and leads to more accurate NBA Finals MVP results (as shown in Table~\ref{table_final}).

\subsubsection{Computational Efficiency of Shapley Values}

We leverage \textbf{TreeSHAP}~\cite{lundberg2020local} to compute feature-level Shapley values directly from the LightGBM model (Section 4.1), which reduces complexity from $\mathcal{O}(TL2^M)$ to $\mathcal{O}(TLD^2)$ (where $T$ = number of trees, $L$ = max leaves, $D$ = depth, $M$ = features) and enables \textbf{linear-time computation} relative to tree size. Our experiments ran on with 4 cores of AMD EPYC 7543 (32-core) and NVIDIA Tesla A30 (24GB VRAM).
\noindent Achieving:
\begin{itemize}
    \item Sub-second response for single-game MVP evaluation
    \item $<$5 minutes for full season NBA analysis
\end{itemize}
Therefore, we can provide MVP evaluation results in real time after a single game ends.

\subsubsection{Online Service Deployment Framework for \oursys}
\begin{figure}[thbp]
  \centering
  \includegraphics[width=\linewidth]{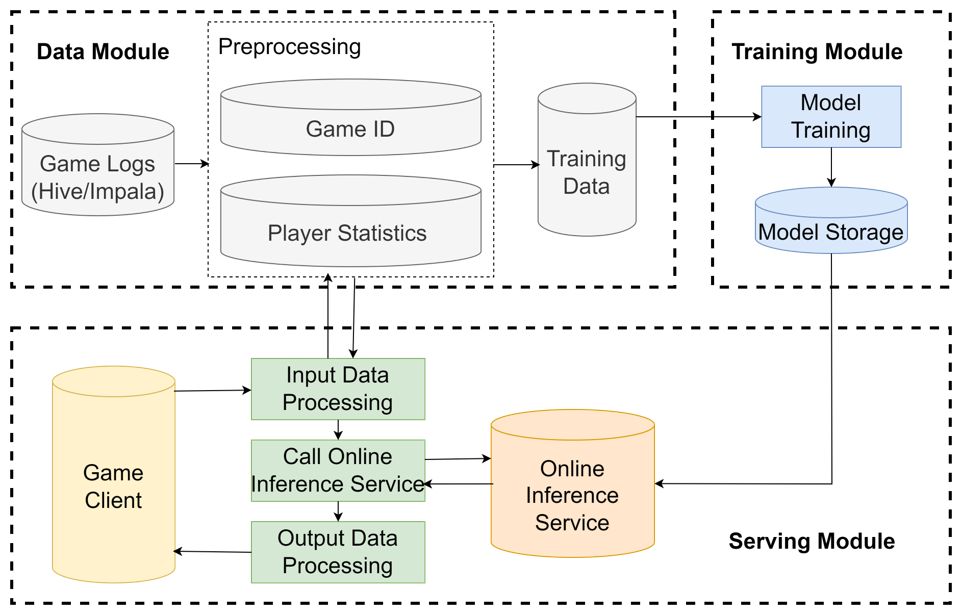}
  \caption{ Online Service Deployment Framework.}
  \label{Deployment}
\end{figure}

The MVP evaluation deployment framework, as shown in Figure~\ref{Deployment}, consists of three main modules: Data, training, and serving.
1) \textbf{Data module:}
Game logs from live basketball games are collected and processed using Hive/Impala. The data, including game ID and player statistics (e.g., point, assist, etc.), is preprocessed to form training data.
2) \textbf{Training module:}
The preprocessed data is used in the model training phase to generate the MVP evaluation model. The trained model is stored in a model storage for later use.
3) \textbf{Serving module:}
Live game data from the game client is processed and passed to an online inference service. Using the pre-trained model, the service evaluates the MVPs in real time, with results sent back to the client for immediate display. This framework provides a scalable and efficient approach to deploying MVP-Shapley, seamlessly integrating historical data with real-time game information to deliver accurate evaluations during live matches.
For the performance comparison with the original method, we have conducted an online A/B test for over one month to compare player report rates ($RR = \frac{N_{\text{reports}}}{N_{\text{players}}} \times 100\%$) and churn rates ($CR = \frac{N_{\text{churned}}}{N_{\text{start\_players}}} \times 100\%$).
Where \( N_{\text{reports}} \) = Number of reports received,
\( N_{\text{players}} \) = Total number of active players during the same period, \( N_{\text{churned}} \) = Number of players who stopped playing the game,
\( N_{\text{start\_players}} \) = Total number of active players at the beginning of the period. 
We employ a rigorous hash-based random bucketing method for our A/B test, with 10,000 players assigned to each bucket.
Our method has reduced the report rate by approximately \textbf{$9.64\text{\textperthousand} \pm 1.2\text{\textperthousand}$} and the churn rate by approximately \textbf{$8.11\text{\textperthousand} \pm 0.9\text{\textperthousand}$} compared to the existing method.
Experimental results indicate the superiority of our method.

%% file: 3_Related_work.tex
\section{Related Work}
Related work can be categorized into three types: Metric Weighting, Machine Learning Techniques, and Cooperative Game Theory among Players.  \textbf{Metric weighting} encompasses a variety of single-metric evaluations. 
For example, PM stands for "Plus-Minus", which measures the difference between the points scored and the points lost by a team when the player is on the court. APM~\cite{madhavan2016predicting,sill2010improved} stands for "Adjusted Plus-Minus", a statistic used to evaluate the impact of a player on the team's score, and RPM~\cite{engelmann2017possession} is a regularized APM.  BPM~\cite{kubatko2007starting,grassetti2021extended} stands for "Box Plus/Minus", which is used to evaluate a player's overall efficiency, considering the player's impact in scoring, rebounding, etc. 
WS stands for "Win Shares", a metric used to measure a player's contribution to his team's wins. WS48~\cite{cao2012sports} stands for "Win Shares per 48 minutes", which considers a player's playing time, making comparisons between different players fairer. 
WARP finds the player’s contribution in terms of how many additional wins he/she brings to the team. VORP~\cite{sarlis2020sports} is a statistic that combines the strengths of BPM and WARP to estimate the points a player contributes per 100 team possessions above that of a replacement-level player.
These metrics can be used to evaluate the MVP individually or in a weighted manner.
The weighting of different metrics was introduced early on~\cite{cooper2009selecting, piette2010scoring}.
\textbf{Machine learning techniques.}
Fearnhead et al.~\cite{fearnhead2011estimating} introduce a new model for evaluating NBA player abilities by comparing team performance with and without the player, controlling for teammates' abilities. The method uses multi-season data to estimate offensive and defensive capabilities, providing an overall player rating.
Page et al.~\cite{page2013effect} model basketball player performance using Gaussian process regression, estimating player performance curves as a function of game percentile.
Metulini et al.~\cite{metulini2020measuring} emphasize evaluating player performances in team sports, estimating scoring probabilities, and developing player-specific shooting performance indices using Classification and Regression Trees (CART) and game data.
Sandri et al.~\cite{sandri2020markov} focus on modeling shooting performance variability and teammate interactions using a Markov switching model, highlighting positive and negative interactions between teammates through network graphs.
Terner et al.~\cite{terner2021modeling} explore various tools for assessing players and discuss the future of basketball analytics, emphasizing the need for causal inference in sports.
\textbf{Cooperative Game Theory among Players}, inspired by previous work~\cite{kolykhalova2020automated,matthiopoulou2020computational}, led Metulini et al.~\cite{metulini2023measuring} to assess players' importance in basketball using the generalized Shapley value. They employed various lineup win-rate prediction models based on historical team roster data to compute the MVP for each team and determine the best lineup.

%% file: 8_Conclusion.tex
\section{Conclusion and Future Work}


In this work, we propose \oursys for evaluating the MVP by leveraging feature-level Shapley values. Our method is both scalable and interpretable with theoretical guarantees. We validated our approach on the NBA and Dunk City Dynasty datasets and successfully deployed it for online industrial use. Future work will focus on enhancing the method’s robustness and stability for broader practical applications. Additionally, we aim to explore finer-grained tracking data, such as player state-action time series, to improve interpretability and uncover deeper insights into player performance.

%% file: 9_appendix.tex
\section{Appendix}

\subsection{Tables}

\begin{table}[htbp]
\centering
\caption{NBA Statistics Summary}
\label{tab:stats_summary}
\begin{tabular}{ll}
\begin{tabularx}{0.9\linewidth}{Xl}
\toprule
\textbf{Basic Statistics} & \textbf{Description} \\
\midrule
MP   & Minutes Played \\
FG   & Field Goals Made \\
FGA  & Field Goals Attempted \\
FG\% & Field Goal Percentage \\
3P   & 3-Point Field Goals Made \\
3PA  & 3-Point Field Goals Attempted \\
3P\% & 3-Point Field Goal Percentage \\
FT   & Free Throws Made \\
FTA  & Free Throws Attempted \\
FT\% & Free Throw Percentage \\
ORB  & Offensive Rebounds \\
DRB  & Defensive Rebounds \\
TRB  & Total Rebounds \\
AST  & Assists \\
STL  & Steals \\
BLK  & Blocks \\
TOV  & Turnovers \\
PF   & Personal Fouls \\
PTS  & Points \\
+/-  & Plus/Minus \\
\bottomrule
\end{tabularx}
\end{tabular}
\quad
\begin{tabular}{ll}
\begin{tabularx}{0.9\linewidth}{Xl}
\toprule
\textbf{Advanced Statistics} & \textbf{Description} \\
\midrule
TS\%   & True Shooting Percentage \\
eFG\%  & Effective Field Goal Percentage \\
3PAr   & 3-Point Attempt Rate \\
FTr    & Free Throw Attempt Rate \\
ORB\%  & Offensive Rebound Percentage \\
DRB\%  & Defensive Rebound Percentage \\
TRB\%  & Total Rebound Percentage \\
AST\%  & Assist Percentage \\
STL\%  & Steal Percentage \\
BLK\%  & Block Percentage \\
TOV\%  & Turnover Percentage \\
USG\%  & Usage Percentage \\
ORtg   & Offensive Rating \\
DRtg   & Defensive Rating \\
BPM    & Box Plus/Minus \\
\bottomrule
\end{tabularx}
\end{tabular}
\end{table}

\clearpage
\subsection{Figures}

\begin{figure}[ht]
    \centering
    \subfigure[]{
        \label{feature_v0} 
        \includegraphics[width=0.47\linewidth, height=7.5cm]
        {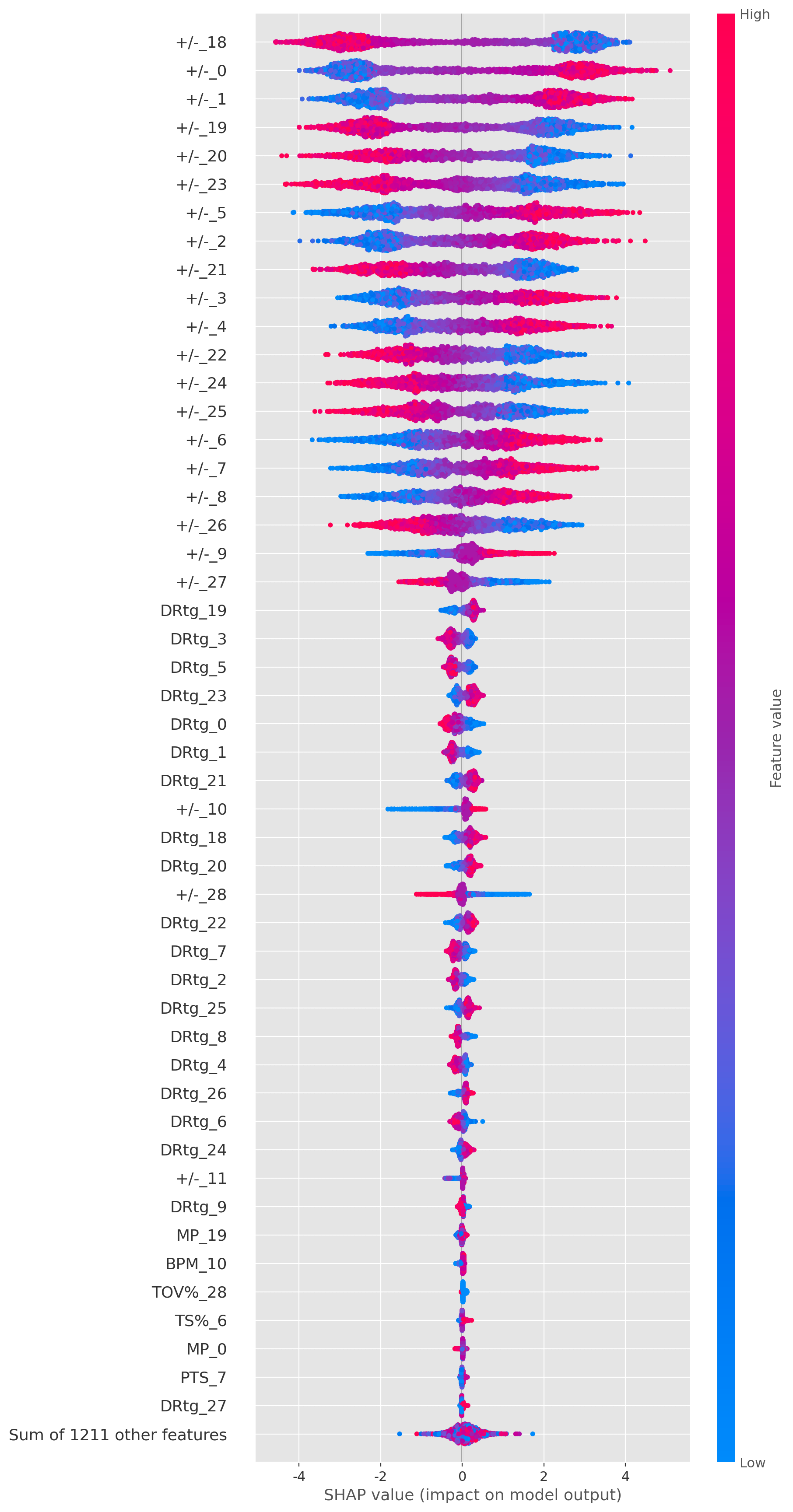}
    }
    \subfigure[]{
        \label{feature_v10} 
        \includegraphics[width=0.47\linewidth, height=7.5cm]
        {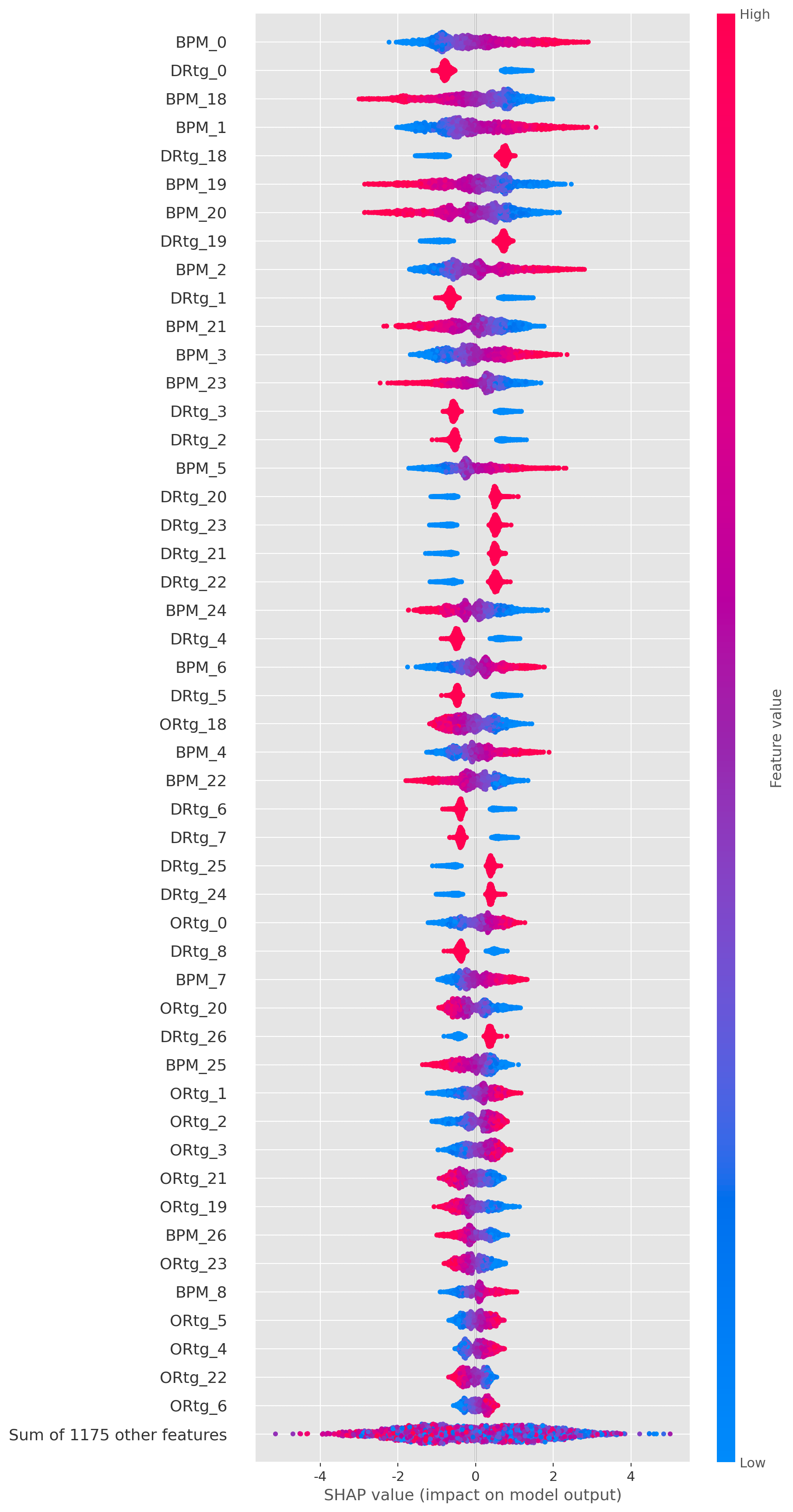}
    }
    
    \caption{NBA dataset. (a) The dot chart visualization of SHAP (SHapley Additive exPlanations) values of all features; (b) The dot chart visualization of SHAP values of the features after fuzzification.}  
    \label{feature_nba} 
\end{figure}

\begin{figure}[ht]
    \centering
    \subfigure[]{
        \label{l33_v1_dots} 
        \includegraphics[width=0.47\linewidth, height=7.5cm]
        {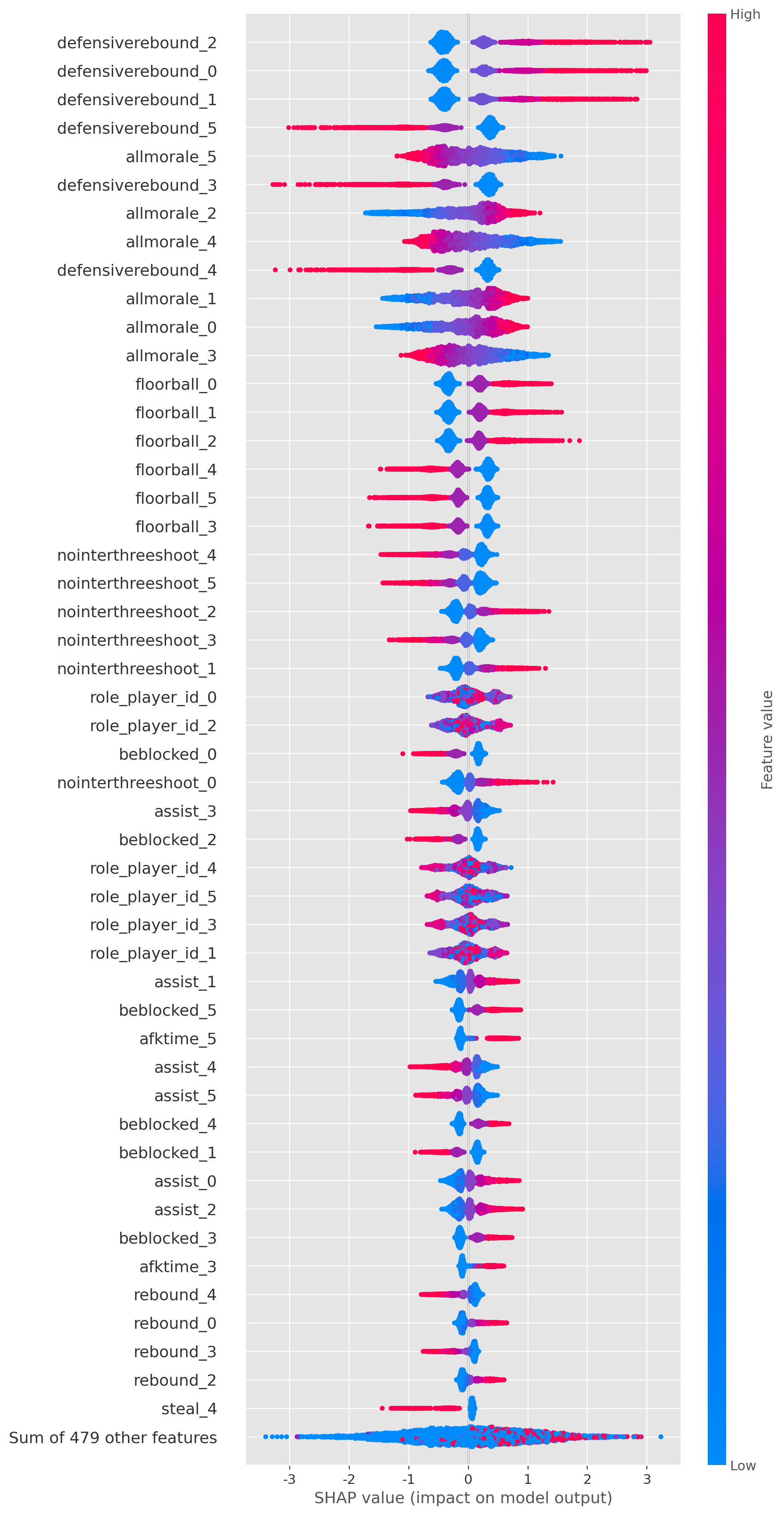}
    }
     \subfigure[]{
        \label{l33_v2_dots} 
        \includegraphics[width=0.47\linewidth, height=7.5cm]
        {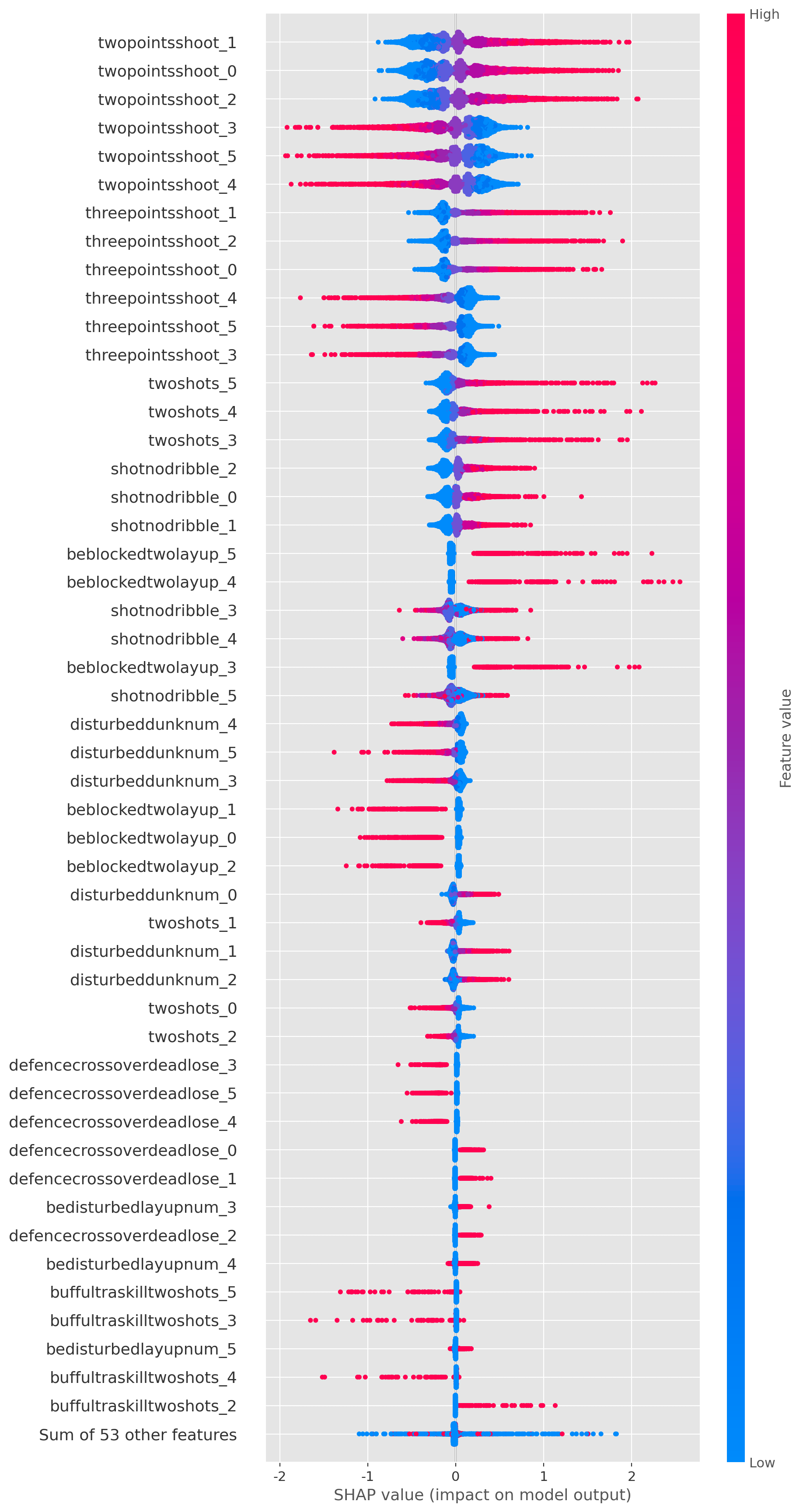}
    }

    \caption{Dunk City Dynasty dataset. (a) The dot chart visualization of SHAP values of all features; (b) The dot chart visualization of SHAP values of optimized features.}  
    \label{feature_l33} 
\end{figure}

\begin{figure}[ht]
    \centering
    \subfigure[]{
        \label{l33_case1} 
        \includegraphics[width=0.47\linewidth, height=7.5cm]{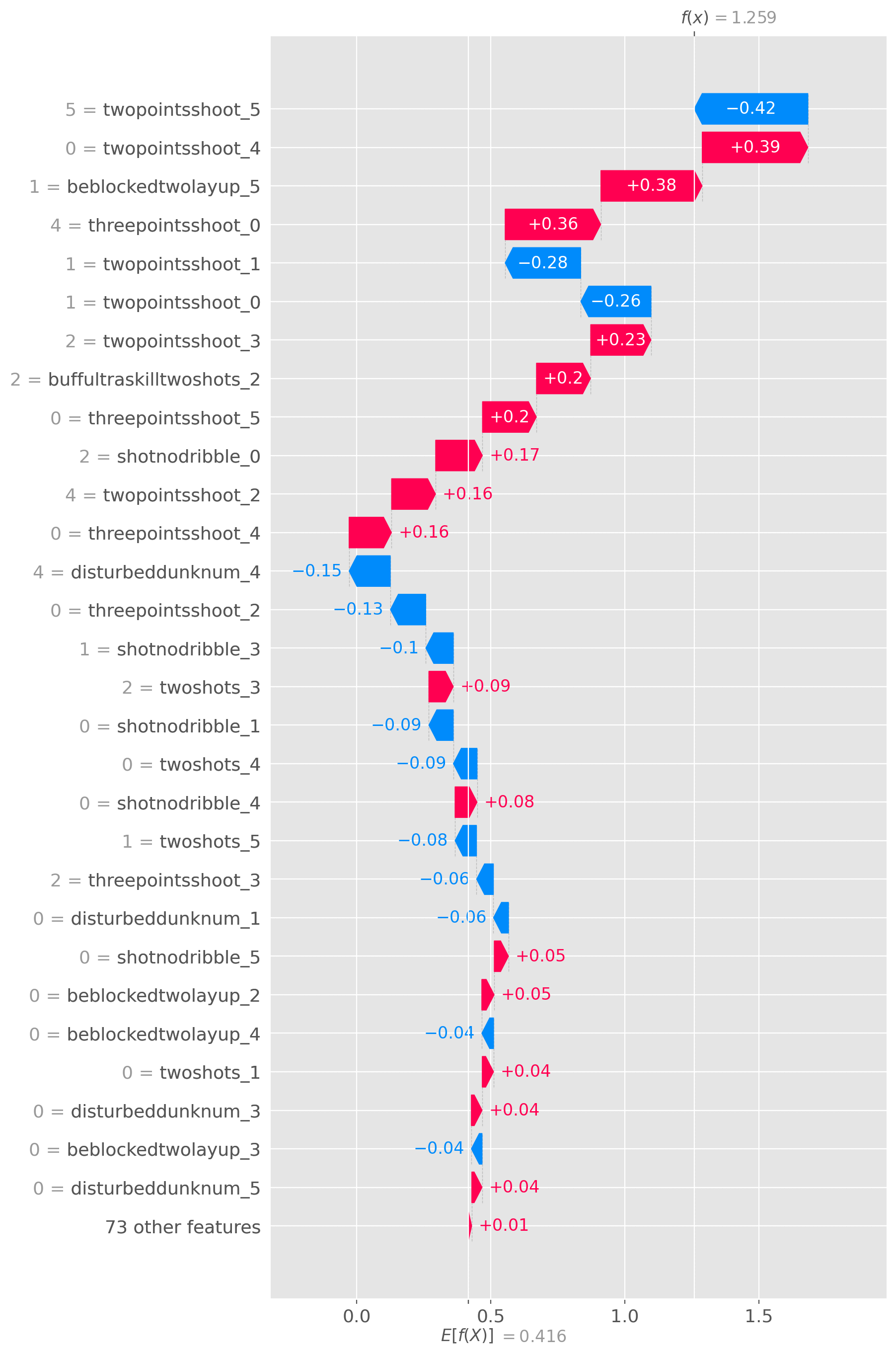}
    }
     \subfigure[]{
        \label{l33_case2} 
        \includegraphics[width=0.47\linewidth, height=7.5cm]{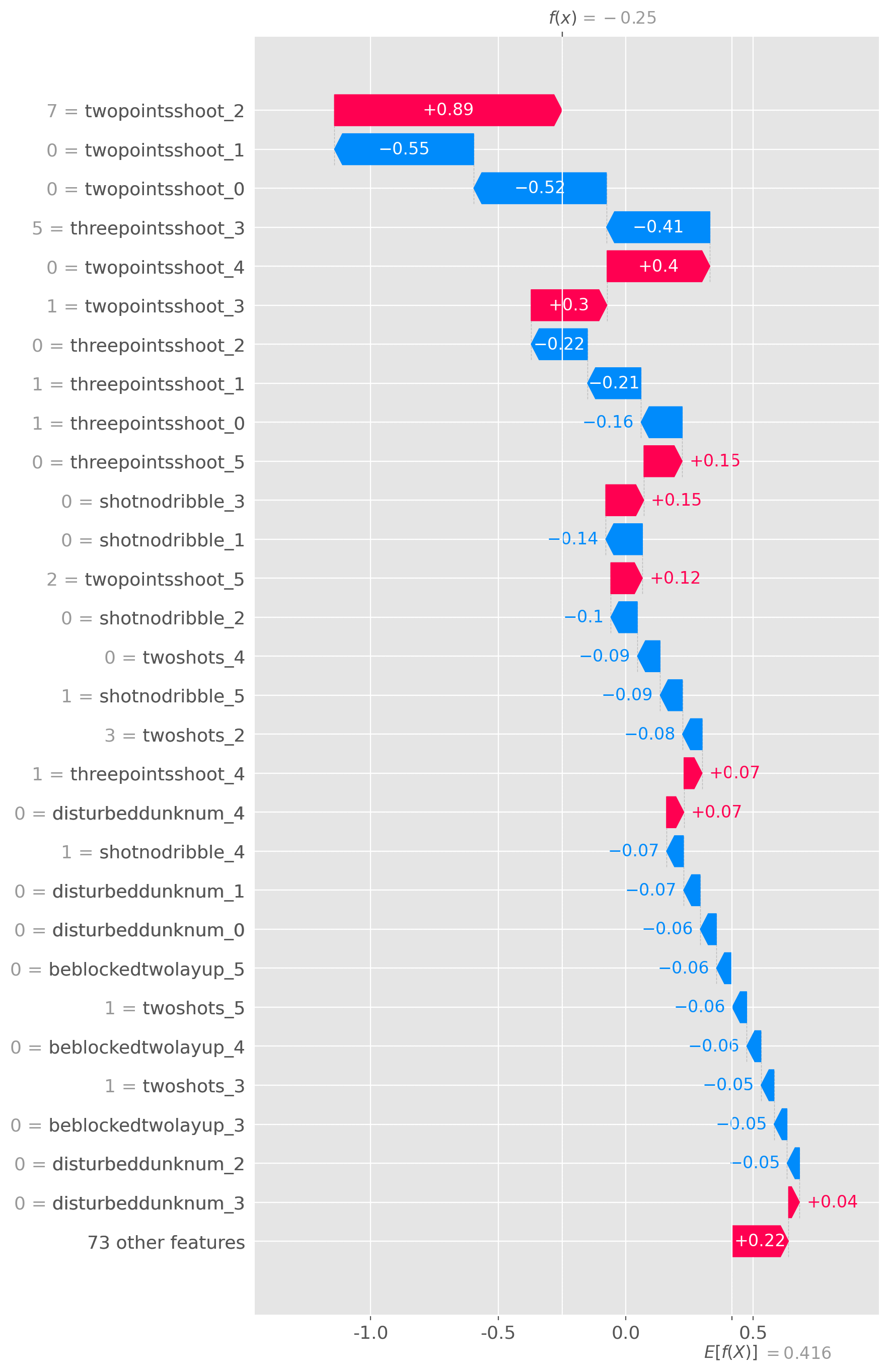}
    }

    \caption{ Dunk City Dynasty dataset. (a) A case analysis of a home team's victory; (b)A case analysis of a home team's loss.}  
    \label{l33_case} 
    \vspace{-0.2in}
\end{figure}

Figure~\ref{feature_nba} and Figure~\ref{feature_l33} respectively demonstrate the change in feature importance before and after optimization for the NBA dataset and the Dunk City Dynasty dataset. Figure~\ref{l33_case} illustrates the impact of different features on the outcome of actual games for Dunk City Dynasty.
Figure~\ref{Deployment} illustrates the deployment framework of our algorithm in the online service. Hive and Impala are both widely used technologies in the Hadoop ecosystem, designed for querying and analyzing large datasets stored in Hadoop Distributed File System (HDFS). 





\subsection{Key Differences Between NBA and Dunk City Dynasty}
\label{Differences}
\subsubsection{Ground Truth Availability}
\begin{itemize}
    \item \textbf{NBA}: Uses \textbf{expert-voted MVP rankings} (real-world ground truth from \texttt{basketball-reference.com})
    \item \textbf{Dunk City Dynasty (DCD)}: As an online game, it lacks intrinsic ground truth. We:
    \begin{itemize}
        \item \textbf{Crowdsourced MVP rankings} via NetEase's platform (Section 4.1.1)
        \item Employed truth inference algorithms (~\cite{dawid1979maximum, weng2024humandata}) and the state-of-the-art voting algorithm, the GOSDT tree~\cite{kang2023learning} to ensure reliability
    \end{itemize}
\end{itemize}

\subsubsection{Game Structure \& MVP Evaluation Scope}
\begin{table}[htbp]
\renewcommand{\arraystretch}{1.5}
\centering
\caption{Structural Comparison Between NBA and DCD}
\begin{tabularx}{\linewidth}{|>{\centering\arraybackslash}p{2.3cm}|>{\centering\arraybackslash}X|>{\centering\arraybackslash}X|}
\hline
\textbf{Feature}         & \textbf{NBA}                & \textbf{DCD} \\ \hline
\textbf{Evaluation Scope} & Multi-game performance & Single-game MVP evaluation \\ \hline
\textbf{Team Composition} & Stable lineups & Random 3v3 matchmaking \\ \hline
\textbf{Data Type} & Real-world play-by-play stats & In-game event logs \\ \hline
\end{tabularx}
\label{NBA and DCD}
\end{table}

The structural comparison between NBA and Dunk City Dynasty (DCD) (Table~\ref{NBA and DCD}) reveals three fundamental differences in their MVP evaluation frameworks:

\textbf{Evaluation Scope}: NBA assesses player performance based on cumulative multi-game statistics over an entire season, while DCD focuses exclusively on single-game MVP determinations.

\textbf{Team Dynamics}: NBA features stable team lineups throughout the season, in contrast to DCD's dynamic random 3v3 matchmaking system where team compositions vary per game.

\textbf{Data Characteristics}: The analysis relies on fundamentally different data sources - traditional play-by-play statistics for NBA versus structured in-game event logs for DCD. This distinction necessitates different methodological approaches for processing and interpreting the respective data types.

These structural differences highlight our framework's adaptability to both traditional sports analytics and esports environments with distinct evaluation requirements.

\subsubsection{Methodological Adaptations}
\begin{itemize}
    \item \textbf{For NBA}:
    \begin{itemize}
        \item Uses traditional basketball stats (e.g., \texttt{DRtg}, \texttt{BPM})
        \item Evaluates season-long contributions
    \end{itemize}
    \item \textbf{For DCD}:
    \begin{itemize}
        \item Processes game-specific metrics (e.g., \texttt{NicePass}, \texttt{Block})
        \item Handles unstable team formations via feature-level Shapley values
    \end{itemize}
\end{itemize}

\subsubsection{Why These Differences Are Strengths}
\begin{enumerate}
    \item \textbf{Generalizability}  
    \begin{itemize}
        \item Handles both:
        \begin{itemize}
            \item Structured, long-term evaluations (NBA)
            \item Dynamic, single-game scenarios (DCD)
        \end{itemize}
    \end{itemize}
    
    \item \textbf{Robustness}  
    \begin{itemize}
        \item Validated on:
        \begin{itemize}
            \item Expert-curated ground truth (NBA)
            \item Crowdsourced consensus (DCD)
        \end{itemize}
    \end{itemize}
    
    \item \textbf{Scalability}  
    \begin{itemize}
        \item Same LightGBM architecture works for:
        \begin{itemize}
            \item Traditional basketball analytics
            \item Esports-specific metrics
        \end{itemize}
    \end{itemize}
\end{enumerate}

\subsection{``Who Voters Pick'' vs. ``True MVP''}
\label{sec:voters-vs-true}

In this paper, we focus on predicting \emph{``who voters pick''} rather than attempting to determine an objective ``true MVP.'' Our goal is to align with expert judgment (voting results), as this represents the practical standard used in professional sports evaluations (e.g., NBA MVP voting).

\begin{itemize}
    \item \textbf{Prediction Target}: Voting outcomes rather than theoretical ideal
    \item \textbf{Alignment Objective}: Match expert voting patterns
    \item \textbf{Practical Basis}: Follows established sports industry standards
\end{itemize}

\section{Proof}
\label{Proof}
\subsection{Axiomatic Properties of Player Contribution}

We first establish that the player contribution measure $\Phi$ inherits the desirable axiomatic properties from Shapley values, adapted to the MVP evaluation context.

\begin{definition}[Player Contribution Measure]
For a game with player sets $\{a_1,\cdots,a_p\}$ and $\{b_1,\cdots,b_p\}$, and constructed data points $x_1, x_2$, where:
\begin{itemize}
    \item In $x_1$: Team A is home, Team B is away
    \item In $x_2$: Team B is home, Team A is away
\end{itemize}
The player contribution measure is defined as:
\begin{equation}
\Phi(a_i,\{x_1,x_2\}) = \sum_{j\in\mathcal{H}(a_i)}\phi_j(x_1)-\sum_{j\in\mathcal{A}(a_i)}\phi_j(x_2),
\end{equation}
where $\phi_j$ denotes the Shapley value of feature $j$, $\mathcal{H}(a_i)$ denotes the feature indices of player $a_i$ when on the home team, and $\mathcal{A}(a_i)$ denotes the feature indices when on the away team.
\end{definition}

\begin{theorem}[Single-Point Efficiency]
\label{thm:efficiency_single}
For any data point $x$, the sum of all feature-level Shapley values equals the model output difference from baseline:
\begin{equation}
\sum_{j=1}^{2pq}\phi_j(x) = f(x)_{win} - f(x_{\emptyset})_{win},
\end{equation}
where $x_{\emptyset}$ represents the baseline prediction.
\end{theorem}

\begin{proof}
This follows directly from the efficiency axiom of Shapley values in cooperative game theory~\cite{shapley1953value}.
\end{proof}

\begin{theorem}[Team Contribution Decomposition]
\label{thm:team_decomp}
For data point $x_1$ where team A is home and team B is away:
\begin{equation}
\sum_{i=1}^{p}\sum_{j\in\mathcal{H}(a_i)}\phi_j(x_1) + \sum_{i=1}^{p}\sum_{j\in\mathcal{A}(b_i)}\phi_j(x_1) = f(x_1)_{win} - f(x_{\emptyset})_{win}.
\end{equation}
Similarly, for $x_2$ where team B is home and team A is away:
\begin{equation}
\sum_{i=1}^{p}\sum_{j\in\mathcal{H}(b_i)}\phi_j(x_2) + \sum_{i=1}^{p}\sum_{j\in\mathcal{A}(a_i)}\phi_j(x_2) = f(x_2)_{win} - f(x_{\emptyset})_{win}.
\end{equation}
\end{theorem}

\begin{proof}
In $x_1$, the feature space is partitioned into home team features (team A, indexed by $\bigcup_{i=1}^{p}\mathcal{H}(a_i)$) and away team features (team B, indexed by $\bigcup_{i=1}^{p}\mathcal{A}(b_i)$). These sets are disjoint and cover all $2pq$ features. By Theorem~\ref{thm:efficiency_single}:
\begin{equation}
\sum_{j=1}^{2pq}\phi_j(x_1) = \sum_{i=1}^{p}\sum_{j\in\mathcal{H}(a_i)}\phi_j(x_1) + \sum_{i=1}^{p}\sum_{j\in\mathcal{A}(b_i)}\phi_j(x_1).
\end{equation}
The proof for $x_2$ follows analogously.
\end{proof}

\begin{theorem}[Relative Contribution Efficiency]
\label{thm:efficiency}
Let $\Phi_H(A, x_1) = \sum_{i=1}^{p}\sum_{j\in\mathcal{H}(a_i)}\phi_j(x_1)$ denote team A's home contribution and $\Phi_A(A, x_2) = \sum_{i=1}^{p}\sum_{j\in\mathcal{A}(a_i)}\phi_j(x_2)$ denote team A's away contribution. Then:
\begin{equation}
\sum_{i=1}^{p}\Phi(a_i,\{x_1,x_2\}) = \Phi_H(A, x_1) - \Phi_A(A, x_2),
\end{equation}
which measures team A's net contribution advantage across both game perspectives.
\end{theorem}

\begin{proof}
By definition of the player contribution measure:
\begin{align}
\sum_{i=1}^{p}\Phi(a_i,\{x_1,x_2\}) &= \sum_{i=1}^{p}\left(\sum_{j\in\mathcal{H}(a_i)}\phi_j(x_1)-\sum_{j\in\mathcal{A}(a_i)}\phi_j(x_2)\right) \nonumber\\
&= \sum_{i=1}^{p}\sum_{j\in\mathcal{H}(a_i)}\phi_j(x_1) - \sum_{i=1}^{p}\sum_{j\in\mathcal{A}(a_i)}\phi_j(x_2) \nonumber\\
&= \Phi_H(A, x_1) - \Phi_A(A, x_2).
\end{align}
\end{proof}

\begin{theorem}[Symmetry]
\label{thm:symmetry}
If two players $a_i$ and $a_j$ have identical feature values in all games, then they have identical contributions:
\begin{equation}
\forall k: a_{ik} = a_{jk} \Rightarrow \Phi(a_i,\{x_1,x_2\}) = \Phi(a_j,\{x_1,x_2\}).
\end{equation}
\end{theorem}

\begin{proof}
By the symmetry property of Shapley values, if features $i$ and $j$ contribute equally to all coalitions, their Shapley values are equal: $\phi_i(x) = \phi_j(x)$.

When $a_{ik} = a_{jk}$ for all $k$, the feature sets $\{a_{i1},\cdots,a_{iq}\}$ and $\{a_{j1},\cdots,a_{jq}\}$ are identical. Therefore, for each corresponding feature index pair $(k_1, k_2)$ where $k_1 \in \mathcal{H}(a_i)$ and $k_2 \in \mathcal{H}(a_j)$:
\begin{equation}
\phi_{k_1}(x_1) = \phi_{k_2}(x_1) \text{ and } \phi_{k_1}(x_2) = \phi_{k_2}(x_2).
\end{equation}

Summing over all features:
\begin{equation}
\sum_{j\in\mathcal{H}(a_i)}\phi_j(x_1) = \sum_{j\in\mathcal{H}(a_j)}\phi_j(x_1),
\end{equation}
and similarly for the away team components, yielding $\Phi(a_i,\{x_1,x_2\}) = \Phi(a_j,\{x_1,x_2\})$.
\end{proof}

\begin{theorem}[Null Player]
\label{thm:null}
A player whose features do not affect the winning probability in either game perspective has zero contribution. Formally, if for all coalitions $S$ and all feature indices $j$ belonging to player $a_i$:
\begin{equation}
f(x_{S\cup\{j\}})_{win} - f(x_S)_{win} = 0, \quad \forall j \in \mathcal{H}(a_i) \cup \mathcal{A}(a_i), \forall S,
\end{equation}
then $\Phi(a_i,\{x_1,x_2\}) = 0$.
\end{theorem}

\begin{proof}
If player $a_i$'s features do not affect the winning probability for any coalition $S$, then by the Shapley value formula:
\begin{equation}
\phi_j(x) = \frac{1}{2pq}\sum_{S\subset N/\{j\}}{\tbinom{2pq-1}{|S|}}^{-1}(f(x_{S\cup\{j\}})_{win} - f(x_S)_{win}) = 0,
\end{equation}
for all $j \in \mathcal{H}(a_i)$ (in data point $x_1$) and all $j \in \mathcal{A}(a_i)$ (in data point $x_2$).

Therefore:
\begin{equation}
\sum_{j\in\mathcal{H}(a_i)}\phi_j(x_1) = 0 \quad \text{and} \quad \sum_{j\in\mathcal{A}(a_i)}\phi_j(x_2) = 0,
\end{equation}
which implies $\Phi(a_i,\{x_1,x_2\}) = \sum_{j\in\mathcal{H}(a_i)}\phi_j(x_1) - \sum_{j\in\mathcal{A}(a_i)}\phi_j(x_2) = 0 - 0 = 0$.
\end{proof}

\subsection{Fairness Guarantees}



\begin{theorem}[Merit-Based Fairness]
\label{thm:Fairness}
The \oursys MVP selection mechanism is merit-based. For a single game, if team A wins:
\begin{equation}
MVP = \arg\max_{a_i}\Phi(a_i,\{x_1,x_2\})
\end{equation}
selects the player with the highest contribution to the victory.
\end{theorem}

\begin{proof}
By construction, $\Phi(a_i,\{x_1,x_2\})$ measures player $a_i$'s marginal contribution to the winning probability. The Shapley value ensures this contribution is computed as the average marginal contribution across all possible player coalitions, providing a fair attribution of the team's success.

Since team A wins, $f(x_1)_{win} > f(x_2)_{win}$, and by Theorem~\ref{thm:efficiency}, the total contribution of team A players is positive. The $\arg\max$ operator selects the player whose features contribute most significantly to this positive outcome, satisfying the merit-based criterion.
\end{proof}

\begin{theorem}[Contribution Additivity Across Games]
\label{thm:additivity}
For multiple games $\{G_1, G_2, \ldots, G_T\}$, the total contribution of player $p$ is additive:
\begin{equation}
\Phi_{total}(p) = \sum_{t=1}^{T}\Phi_t(p),
\end{equation}
where $\Phi_t(p)$ is the contribution in game $t$.
\end{theorem}

\begin{proof}
Each game $G_t$ is an independent event with its own data points $\{x_1^{(t)}, x_2^{(t)}\}$. The Shapley value computation for each game is performed independently on the respective win-loss model. Since the contribution measure $\Phi$ is defined per game, and there are no cross-game feature interactions in our model:
\begin{equation}
\Phi_t(p) = \sum_{j\in\mathcal{H}_t(p)}\phi_j(x_1^{(t)})-\sum_{j\in\mathcal{A}_t(p)}\phi_j(x_2^{(t)})
\end{equation}
is computed independently for each $t$.

The total contribution is simply the sum of individual game contributions, satisfying additivity.
\end{proof}

\subsection{Consistency of Multi-Game MVP Evaluation}

We analyze the consistency properties of the MVP evaluation methods proposed for multiple games.

\begin{theorem}[Ranking Consistency of $M_1$]
\label{thm:m1}
For Method $M_1$ (average ranking in winning games), let $R_i^{(t)}$ denote the rank of player $i$ in game $t$. If player $i$ consistently ranks higher than player $j$ in all winning games:
\begin{equation}
\forall t \in G_{win}: R_i^{(t)} < R_j^{(t)} \Rightarrow \frac{1}{|G_{win}|}\sum_{t\in G_{win}}R_i^{(t)} < \frac{1}{|G_{win}|}\sum_{t\in G_{win}}R_j^{(t)}.
\end{equation}
\end{theorem}

\begin{proof}
If $R_i^{(t)} < R_j^{(t)}$ for all $t \in G_{win}$, then:
\begin{equation}
\sum_{t\in G_{win}}R_i^{(t)} < \sum_{t\in G_{win}}R_j^{(t)}.
\end{equation}
Dividing both sides by $|G_{win}|$ (which is positive) preserves the inequality:
\begin{equation}
\frac{1}{|G_{win}|}\sum_{t\in G_{win}}R_i^{(t)} < \frac{1}{|G_{win}|}\sum_{t\in G_{win}}R_j^{(t)}.
\end{equation}
Thus, $M_1$ respects consistent ranking dominance.
\end{proof}

\begin{theorem}[Value Consistency of $M_3$]
\label{thm:m3}
For Method $M_3$ (average contribution), if player $i$ has higher contribution than player $j$ in every game:
\begin{equation}
\forall t: \Phi_t(i) > \Phi_t(j) \Rightarrow \frac{1}{T}\sum_{t=1}^{T}\Phi_t(i) > \frac{1}{T}\sum_{t=1}^{T}\Phi_t(j).
\end{equation}
\end{theorem}

\begin{proof}
If $\Phi_t(i) > \Phi_t(j)$ for all $t \in \{1,\ldots,T\}$, then:
\begin{equation}
\sum_{t=1}^{T}\Phi_t(i) > \sum_{t=1}^{T}\Phi_t(j).
\end{equation}
Dividing by $T > 0$:
\begin{equation}
\frac{1}{T}\sum_{t=1}^{T}\Phi_t(i) > \frac{1}{T}\sum_{t=1}^{T}\Phi_t(j).
\end{equation}
\end{proof}
\subsection{Uniqueness of Shapley-Based MVP Evaluation}


\begin{theorem}[Uniqueness Theorem]
\label{thm:unique}
The \oursys player contribution measure $\Phi$ is the unique attribution method at the feature level satisfying:
\begin{enumerate}
    \item \textbf{Efficiency} (Theorem~\ref{thm:efficiency_single})
    \item \textbf{Symmetry} (Theorem~\ref{thm:symmetry})
    \item \textbf{Null Player} (Theorem~\ref{thm:null})
    \item \textbf{Additivity} (Theorem~\ref{thm:additivity})
\end{enumerate}
\end{theorem}

\begin{proof}
This follows from the Shapley uniqueness theorem~\cite{shapley1953value}. The Shapley value is the unique value function satisfying efficiency, symmetry, dummy player, and additivity axioms. 

At the feature level, for each data point $x$, the Shapley values $\{\phi_j(x)\}_{j=1}^{2pq}$ are uniquely determined by these axioms. Since our player contribution measure $\Phi$ is derived by aggregating feature-level Shapley values:
\begin{equation}
\Phi(a_i,\{x_1,x_2\}) = \sum_{j\in\mathcal{H}(a_i)}\phi_j(x_1)-\sum_{j\in\mathcal{A}(a_i)}\phi_j(x_2),
\end{equation}
the uniqueness of $\phi_j$ implies the uniqueness of $\Phi$ given the aggregation scheme.

Suppose there exists another attribution method $\Phi'$ with a different feature-level attribution $\phi'$. If $\phi'$ satisfies the four Shapley axioms, then $\phi'_j = \phi_j$ by the uniqueness theorem, and consequently $\Phi' = \Phi$.
\end{proof}

\subsection{Convergence Analysis}


\begin{theorem}[Law of Large Numbers for MVP Ranking]
\label{thm:lln}
Let $\mu_i = \mathbb{E}[\Phi(p_i)]$ be the expected contribution of player $p_i$. For method $M_3$, as the number of games $T \rightarrow \infty$:
\begin{equation}
\bar{\Phi}_T(p_i) = \frac{1}{T}\sum_{t=1}^{T}\Phi_t(p_i) \xrightarrow{a.s.} \mu_i,
\end{equation}
where $\xrightarrow{a.s.}$ denotes almost sure convergence.
\end{theorem}

\begin{proof}
Assuming the game contributions $\{\Phi_t(p_i)\}_{t=1}^{T}$ are independent and identically distributed (i.i.d.) with finite mean $\mu_i$ and variance $\sigma_i^2$, the Strong Law of Large Numbers applies:
\begin{equation}
P\left(\lim_{T\rightarrow\infty}\bar{\Phi}_T(p_i) = \mu_i\right) = 1.
\end{equation}

While the i.i.d. assumption may not hold perfectly in practice (e.g., player form varies over time), the ergodic theorem provides similar guarantees under stationary conditions.
\end{proof}

\begin{theorem}[Concentration Bound]
\label{thm:concentration}
For any player $p_i$ with bounded contribution $|\Phi_t(p_i)| \leq M$, with probability at least $1-\delta$:
\begin{equation}
\left|\bar{\Phi}_T(p_i) - \mu_i\right| \leq M\sqrt{\frac{2\ln(2/\delta)}{T}}.
\end{equation}
\end{theorem}

\begin{proof}
By Hoeffding's inequality, for bounded random variables in $[-M, M]$:
\begin{equation}
P\left(\left|\bar{\Phi}_T(p_i) - \mu_i\right| \geq \epsilon\right) \leq 2\exp\left(-\frac{T\epsilon^2}{2M^2}\right).
\end{equation}

Setting the right-hand side equal to $\delta$ and solving for $\epsilon$:
\begin{equation}
\epsilon = M\sqrt{\frac{2\ln(2/\delta)}{T}}.
\end{equation}
\end{proof}

\begin{corollary}[Sample Complexity for MVP Identification]
\label{cor:sample}
To distinguish between two players with expected contribution difference $\Delta = |\mu_i - \mu_j|$ with probability at least $1-\delta$, the required number of games is:
\begin{equation}
T \geq \frac{8M^2\ln(4/\delta)}{\Delta^2}.
\end{equation}
\end{corollary}

\begin{proof}
Applying Theorem~\ref{thm:concentration} to both players and using a union bound:
\begin{equation}
P\left(\left|\bar{\Phi}_T(p_i) - \mu_i\right| < \frac{\Delta}{2} \text{ and } \left|\bar{\Phi}_T(p_j) - \mu_j\right| < \frac{\Delta}{2}\right) \geq 1-\delta.
\end{equation}

When both events hold, if $\mu_i > \mu_j$, then:
\begin{equation}
\bar{\Phi}_T(p_i) > \mu_i - \frac{\Delta}{2} = \mu_j + \frac{\Delta}{2} > \bar{\Phi}_T(p_j).
\end{equation}

Setting $\epsilon = \Delta/2$ in Hoeffding's bound with $\delta/2$ for each player:
\begin{equation}
T \geq \frac{2M^2\ln(4/\delta)}{(\Delta/2)^2} = \frac{8M^2\ln(4/\delta)}{\Delta^2}.
\end{equation}
\end{proof}

\subsection{Theoretical Analysis of Fuzzification}


\begin{definition}[Confounding Variable]
A variable $X_c$ is a \textit{confounding variable} if:
\begin{enumerate}
    \item $X_c$ causally influences the outcome $Y$ (game result).
    \item $X_c$ causally influences other variables $X_i$ used in the evaluation.
\end{enumerate}
\end{definition}

\begin{theorem}[Bias Reduction through Fuzzification]
\label{thm:fuzz}
Let $X_c$ be a confounding variable with range $[a,b]$. The fuzzified variable $\hat{X}_c$ with $t$ bins reduces the mutual information between $\hat{X}_c$ and other variables:
\begin{equation}
I(\hat{X}_c; X_i) \leq I(X_c; X_i),
\end{equation}
where $I(\cdot;\cdot)$ denotes mutual information.
\end{theorem}

\begin{proof}
Fuzzification is a deterministic function $g: X_c \rightarrow \hat{X}_c$ that maps continuous values to discrete bins. By the data processing inequality:
\begin{equation}
I(\hat{X}_c; X_i) = I(g(X_c); X_i) \leq I(X_c; X_i).
\end{equation}

The equality holds only when $g$ is a bijection, which is not the case for fuzzification with $t < \infty$ bins. Therefore, fuzzification strictly reduces the mutual information when $t$ is finite, reducing the confounding effect.
\end{proof}

\begin{corollary}[Optimal Bin Size]
\label{cor:bin}
There exists an optimal bin size $t^*$ that balances bias reduction and information preservation:
\begin{equation}
t^* = \arg\min_t \left[\lambda \cdot \text{Bias}(t) + (1-\lambda) \cdot \text{Var}(t)\right],
\end{equation}
where $\lambda \in [0,1]$ is a trade-off parameter.
\end{corollary}